\documentclass[aps,prl,superscriptaddress,notitlepage,nopacs,amsmath,amstex,amssymb,citeautoscript,longbibliography,floatfix,letter,twocolumn]{revtex4-2}
\usepackage{graphicx}
\usepackage{subfigure}
\usepackage{adjustbox}
\usepackage{bm}
\usepackage{color}
\usepackage{braket}
\usepackage{standalone}
\usepackage{multirow}
\usepackage{tikz}
\usepackage{mathrsfs}
\usepackage{dsfont}
\usepackage{comment}
\usepackage{amsmath}
\usepackage[colorlinks,bookmarks=true,citecolor=blue,linkcolor=red,urlcolor=blue]{hyperref}
\usepackage{cleveref}
\usepackage{orcidlink}

\renewcommand{\vec}[1]{\mbox{\boldmath$#1$}}

\graphicspath{{./figures/}}

\begin{document}

\title{Splitting of Girvin-MacDonald-Platzman density wave and the nature of chiral gravitons in fractional quantum Hall effect}

\author{Ajit C. Balram\orcidlink{0000-0002-8087-6015}}
\email{cb.ajit@gmail.com}
\affiliation{Institute of Mathematical Sciences, CIT Campus, Chennai, 600113, India}
\affiliation{Homi Bhabha National Institute, Training School Complex, Anushaktinagar, Mumbai 400094, India}

\author{G. J. Sreejith\orcidlink{0000-0002-2068-1670}}
\email{sreejith@acads.iiserpune.ac.in}
\affiliation{Indian Institute of Science Education and Research, Pune, India 411008}

\author{J. K. Jain\orcidlink{000-0003-0082-5881}}
\email{jkj2@psu.edu}
\affiliation{Department of Physics, 104 Davey Lab, Pennsylvania State University, University Park, Pennsylvania 16802, USA}

\date{\today}

\begin{abstract}
A fundamental manifestation of the nontrivial correlations of an incompressible fractional quantum Hall (FQH) state is that an electron added to it disintegrates into more elementary particles, namely fractionally-charged composite fermions (CFs). We show here that the Girvin-MacDonald-Platzman (GMP) density-wave excitation of the $\nu{=}n/(2pn{\pm }1)$ FQH states also splits into more elementary single CF excitons. In particular, the GMP graviton, which refers to the recently observed spin-2 neutral excitation in the vanishing wave vector limit [Liang {\it et al.}, Nature {\bf 628}, 78 (2024)], remains undivided for $\nu{=}n/(2n{\pm} 1)$ but splits into two gravitons at $\nu{=}n/(4n{\pm} 1)$ with $n{>}1$. A detailed experimental confirmation of the many observable consequences of the splitting of the GMP mode should provide a unique window into the correlations underlying the FQH effect. 
\end{abstract}

\maketitle

\textbf{\textit{Introduction.}} 
A dramatic manifestation of strong correlations is ``fractionalization of the electron," which refers to the notion that the emergent elementary particles (that is, the weakly interacting degrees of freedom) carry quantum numbers of a split electron~\cite{Anderson97}. For example, when an electron is added to a Luttinger liquid, its spin and charge separate~\cite{Giamarchi04}, and when a fully spin-polarized electron is added to an incompressible FQH state, it disintegrates into fractionally-charged excited CFs~\cite{Jain07, Halperin20}. The emergent particles have their own independent dynamics and provide a natural explanation of the otherwise puzzling phenomenology.  

This Letter concerns the Girvin-MacDonald-Platzman (GMP) density wave mode of the FQH states~\cite{Girvin85, Girvin86}. Since the work of GMP numerous experimental~\cite{Pinczuk93, Mellor95, Kang00, Kang01, Dujovne03, Hirjibehedin03, Hirjibehedin05, Gallais06, Kukushkin07, Kukushkin09, Rhone11, Kukushkin09} and theoretical~\cite{Dev92, Lopez93, Simon93, Simon94a, Scarola00, Park00, Ghosh01, Majumder09, Haldane11, Sreejith11, Yang12b, Yang13a,  Yang14, Majumder14, Golkar16, Golkar16a, Balram16d, Nguyen17b, Nguyen18, Balram21d, Nguyen22, Wang22} studies have addressed the neutral excitations of the FQH states. Very recent inelastic light scattering experiments by Liang {\it et al.}~\cite{Liang24} with circularly polarized photons have measured the chirality of the long wavelength neutral excitation, which has been called ``graviton" because of its spin-2 quadrupolar character in a geometric interpretation, and found it to be consistent with theoretical predictions~\cite{Haldane11, Golkar16, Gromov17}. 

We show in this Letter that the GMP density wave mode also fragments into more elementary CF excitons (CFEs), which suggests a generalization of GMP's single mode approximation (SMA) [Eq.~\eqref{eq: SMA}] for the dynamical structure factor (DSF) to a ``CFE ansatz" [Eq.~\eqref{eq: CFA}]. We further show that while, surprisingly, the GMP and the CF theories provide essentially identical descriptions for the graviton of the $n/(2n{\pm} 1)$ states, the GMP graviton breaks into two elementary gravitons for the $n/(4n{\pm} 1)$ states with $n{>}1$. We discuss the experimental implications of the GMP mode splitting.

\textbf{\textit{SMA.}} 
Let us begin by recalling the GMP ansatz. The lowest Landau level (LLL)-projected DSF at any filling factor $\nu$ is given by (modulo an overall factor)
\begin{equation} 
\label{eq: Sqomega}
S(\vec{q},E)=
\sum_\alpha |\langle \Psi_{\nu}^\alpha|  \Psi^{\rm GMP}_{\vec{q}}\rangle|^2 \delta(E_\alpha-E),
\end{equation}
where $\Psi_{\nu}^\alpha$ are the LLL many-particle eigenstates with energy $E_\alpha$, $\Psi^{\rm GMP}_{\vec{q}}{=}{\cal P}_{\rm LLL}\rho_{_{\vec{q}}}\Psi_{\nu}^0{=}\bar{\rho}_{\vec{q}}\Psi_{\nu}^0$, and $\Psi_{\nu}^0$ is the LLL ground state. The projected density operator is defined as $\bar{\rho}_{_{\vec{q}}}{=}{\cal P}_{\rm LLL}\rho_{\vec{q}} {\cal P}_{\rm LLL}$, where $\rho_{_{\vec{q}}}{=}\sum_{\vec{k}}c^{\dagger}_{\vec{k}+\vec{q}}c_{\vec{k}}$ creates an electron-hole pair with wave vector $\vec{q}$ and ${\cal P}_{\rm LLL}$ implements projection to LLL. GMP proceeded by making an SMA\cite{Girvin85, Girvin86} which assumes that the GMP mode $\Psi^{\rm GMP}_{\vec{q}}$
has overlap with a single eigenstate, leading to
\begin{equation}\label{eq: SMA}
    S(\vec{q}, E){=}S_{\vec{q}}\delta(E{-}E^{\rm SMA}_{\vec{q}}).
\end{equation} 
Here $S_{\vec{q}}$ is the LLL-projected static structure factor. In general, $E^{\rm SMA}_{\vec{q}}$ is the ``average energy" of the excitations coupled to the ground state by $\bar{\rho}_{_{\vec{q}}}$. $\Psi^{\rm GMP}_{\vec{q}}$ can be evaluated explicitly for small systems for which the Fock space representation of the exact or the trial ground state is available but not for larger systems. GMP related the energy of this mode to the unprojected static structure factor, which is the Fourier transform of the pair correlation function $g(\vec{r})$, which can be evaluated for a trial wave function for fairly large systems~\cite{Park96, Kamilla97, Balram15b, Balram17}, allowing for a determination of the energy of the GMP mode~\cite{Girvin85, Girvin86, Park00a, Scarola00}. 

\begin{figure}[t]
		\centering    
		\includegraphics[width=\linewidth]{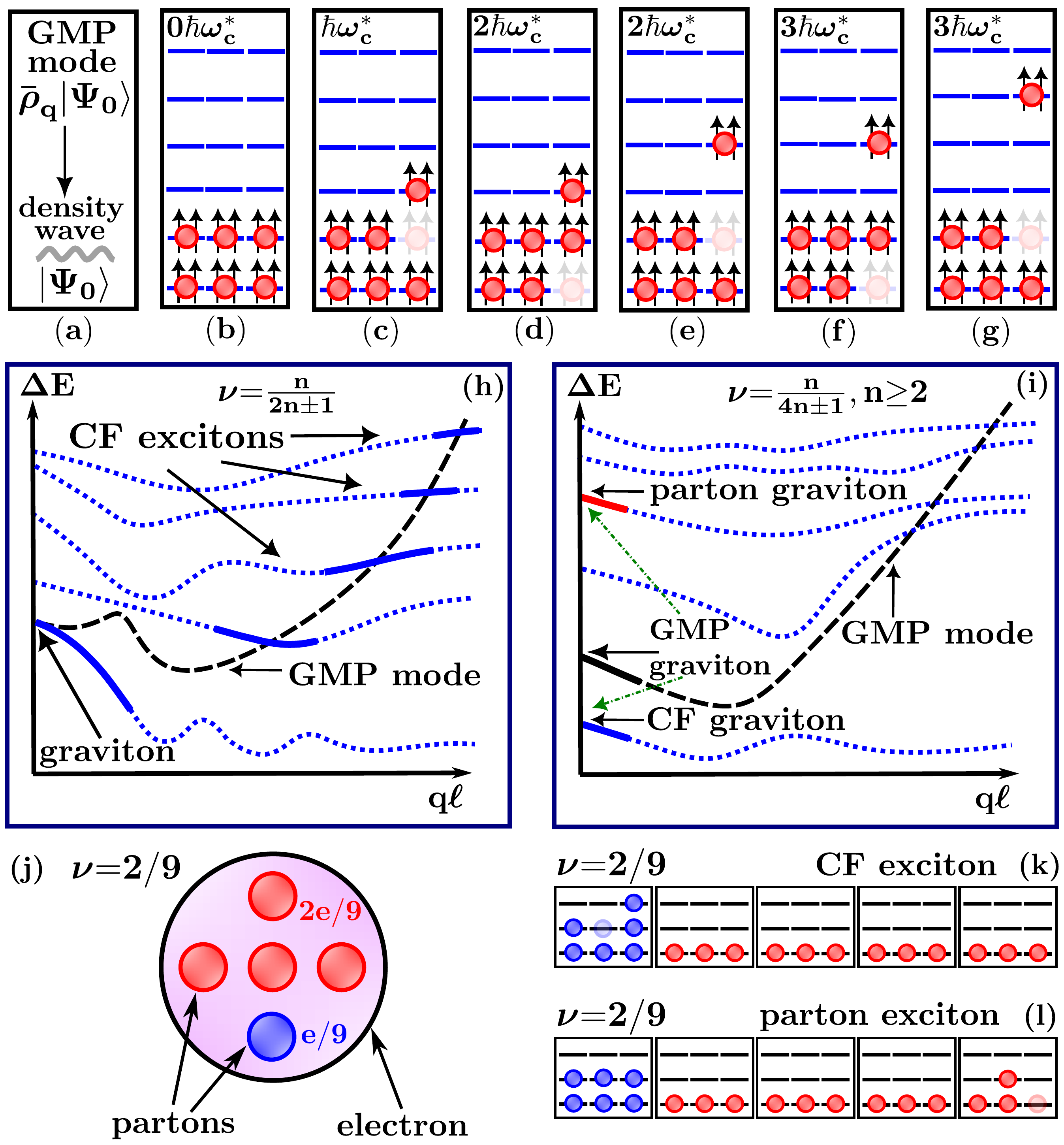}
		\caption{(a) GMP density wave mode, created by applying LLL-projected electron density operator on the ground state. (b) Jain CF ground state at $\nu{=}2/5$. (c)-(g) Examples of single CF excitons. (h),(i) Schematic depiction of splitting of the GMP density wave mode (dashed black curve) into a ladder of CF excitons (blue dotted) which can be obtained by diagonalizing the Coulomb interaction in the CFE basis. The actual lowest neutral mode is well described as the primary CFE shown in (c). For $\nu{=}n/(2n{\pm} 1)$ the GMP graviton is essentially identical to the primary CF graviton (h) while for $\nu{=}n/(4n{\pm} 1)$ with $n{>}1$ the GMP graviton splits into two CF gravitons, which can be also be nicely understood as the primary CF and parton gravitons (i). (j) Parton construction of $\nu{=}2/9$. Each electron is divided into one parton of charge $e/9$ filling two LLs and four partons of charge $2e/9$, each filling a single LL. (k) CF exciton. (l) Parton exciton. }
		\label{fig: schematic}
	\end{figure}

\textbf{\textit{CF exciton.}} 
The CF physics of the FQH states~\cite{Jain89} suggests a different structure for the excitations. The Jain wave function for the ground state of $N$ electrons at $\nu{=}n/(2pn{\pm} 1)$ is given by $\Psi_{n/(2pn\pm 1)}^0{=}{\cal P}_{\rm LLL} \Phi_{\pm n} \Phi_1^{2p}$, where $\Phi_n$ is the wave function of $n$ filled LLs with $\Phi_{{-}n}{=}\Phi_{n}^*$ and $\Phi_1{\sim} \prod_{1{\leq}j{<}k{\leq}N} (z_j{-}z_k)$ with $z_j{=}x_j{-}iy_j$ denoting the position of the $j^{\rm th}$ electron. 
$\Psi_{\nu}^0$ is interpreted as $n$ filled CF LLs (called $\Lambda$Ls) and the neutral excitations as CFEs, namely a particle-hole pair of CFs, as shown in Fig.~\ref{fig: schematic}. The wave function of a single CFE across $K$ $\Lambda$Ls ($K{=}1,2,{\cdots}$) is given by~\cite{Dev92, Kamilla96b, Kamilla96c, Scarola00}
\begin{equation}
	\label{eq: CFE}
	\Psi^{{\rm CF-ex}}_{\vec{q},m\rightarrow m+K}={\cal P}_{\rm LLL}[\rho_{_{\vec{q}}}^{m\rightarrow m+K} \Phi_n] \Phi_1^{2p},
\end{equation}
where $\rho_{_{\vec{q}}}^{m\rightarrow m+K}$ creates an electron-hole pair with wave vector $\vec{q}$ in $\Phi_n$, with hole in the LL with index $m{=}0,1,{\cdots},n{-}1$ and electron in LL with index $m{+}K{=}n,n{+}1,{\cdots}$. Multiplication by $\Phi_1^{2p}$ converts this into a CF-particle CF-hole pair with wave vector $\vec{q}$.

We employ two methods for treating the LLL projection. In ``direct projection" one expresses the wave function in the Fock basis and retains only the part that resides in the LLL~\cite{Dev92}. This method can be implemented only for small systems, as the dimension of the Fock space grows exponentially with $N$ [see Supplemental Material (SM)~\cite{SM_Balram_Sreejith_Jain_Graviton}]. The Jain-Kamilla (JK) method~\cite{Jain97, Jain97b, Moller05, Davenport12, Balram15a} allows treatment of much larger systems. The two methods generally produce wave functions that are extremely close but not identical.

\textbf{\textit{Spherical geometry.}} 
We use the spherical geometry~\cite{Haldane83} in which $N$ electrons move on the surface of a sphere in the presence of a radial magnetic field generated by a magnetic monopole producing a total flux $2Qhc/e$, where $2Q$ is an integer. The LL orbitals are given by the monopole harmonics $Y_{Q,l,l_z}$~\cite{Wu76, Wu77}, where $l{=}|Q|, |Q|{+}1, {\cdots} $ and $l_z{=}{-}l, {-}l{+}1, {\cdots}, l$. The $n$th LL corresponds to $l{=}|Q|{+}n$, with $n{=}0,1,2{\cdots}$. The eigenstates are labeled by the total orbital angular momentum $L$ and its $z$-component $L_z$. 

In first quantized notation, the density operator is given by $\rho_{L, L_{z}}{=}\sum_{i{=}1}^{N}Y_{Q, L,L_{z}}(u_{i}, v_{i})$, where $u_{i}{=}\cos(\theta_{i}/2)e^{i\phi_{i}/2},~v_{i}{=}\sin(\theta_{i}/2)e^{{-}i\phi_{i}/2}$ are the $i^{\rm th}$ electron's spinor coordinates~\cite{He94}. The GMP mode with $L{=}0$ is identical to the ground state $\Psi^0_{\nu}$ and the $L{=}1$ GMP mode is annihilated upon projection to the LLL~\cite{He94}. The $L{=}2$ GMP mode is called the spin-2 graviton. The GMP mode can be constructed up to $L^{\rm GMP}_{\rm max}{=}2Q{\sim}N/\nu$ for large $N$ because the GMP operator creates an {\it electron} and a {\it hole} in the LLL each of which carries an angular momentum of $l{=}Q$. The wave number is given by $|\vec{q}|{=}q{=}L/(\sqrt{Q}\;\ell)$, where $\ell{=}\sqrt{\hbar c/(eB)}$ is the magnetic length at field $B$. 

For CFE in Eq.~\eqref{eq: CFE}, $\Phi_{n}$ is the wave function of $n$ filled LLs of electrons at an effective flux of $2Q^{*}{=}2Q{-}2(N{-}1){=}N/n{-}n$. There are many CFE modes corresponding to excitations in $\Phi_n$ across different LLs. The lowest energy CFE, called the ``primary CFE" [Fig.~\ref{fig: schematic}(c)], corresponds to $n{-}1{\rightarrow} n$ excitation. The angular momenta of the constituent CF particle and CF hole are given by $l{=}Q^{*}{+}n$ and $l{=}Q^{*}{+}n{-}1$. The primary CFE extends from $L^{\rm CFE}_{\rm min}{=}2$ (the $L{=}1$ CFE is also annihilated upon projection into the LLL~\cite{Dev92}) to $L^{\rm CFE}_{\rm max}{=}(N/n{+}n{-}1){\sim}N/n{=}N/\nu^{*}$ for large $N$, where $\nu^{*}$ is the effective filling of CFs~\cite{Balram16d}. From a naive counting, at $L{=}2$, there are two CFEs for $\nu{=}1/(2p{+}1)$ ($0{\rightarrow}1$ and $0{\rightarrow} 2$ excitons), and three for $\nu{=}n/(2pn{+}1)$ for $n{\geq} 2$ ($n{-}1{\rightarrow} n$, $n{-}1{\rightarrow} n{+}1$ and $n{-}2{\rightarrow}n$); for $L{=}3$ we have three CFEs for $\nu{=}1/(2p{+}1)$, five for $\nu{=}2/(4p{+}1)$ and 6 for $\nu{=}n/(2pn{+}1)$ for $n{\geq} 3$; and so on.

\textbf{\textit{Splitting of GMP mode.}} 
Our numerical studies (below) strongly suggest that the GMP density wave lives fully in the subspace defined by the single CFEs, i.e.:
\begin{equation}
	\label{conjecture1}
\Psi^{\rm GMP}_{\vec{q}}\equiv \bar{\rho}_{\vec{q}}\Psi^0= \sum_{\alpha} C_{\alpha} \Psi^{{\rm CFE}}_{\vec{q},\alpha},
\end{equation}
where $\{\Psi^{{\rm CFE}}_{\vec{q},\alpha} \}$ is an orthonormal basis constructed from 
$\{\Psi^{{\rm CF-ex}}_{\vec{q},m\rightarrow m{+}K}\}$.  

We construct the Fock space representations of the direct and JK-projected Jain wave functions for the ground and CFE states by the method given in Refs.~\cite{Sreejith13, Balram20a} which involves an expansion of the state in the set of all the relevant $L$ eigenstates. We construct an orthonormal basis for the linearly independent CFEs and denote by ${\cal O}{=}\sum_{\alpha}|C_{\alpha}|^2$ the sum of squared overlaps of the GMP mode with the orthonormalized basis functions. Eq.~\eqref{conjecture1} is equivalent to ${\cal O}{=}1$.

In all cases studied, the GMP mode and CFEs constructed with direct or JK projection vanish identically for $L{=}1$. For higher $L$, we have studied the following ($\nu$, $N$) systems with direct projection: ($1/3$,6) for $L{\leq }7$; ($1/3$,8) for $L{\leq}7$; ($2/5$,6) for $L{\leq}4$; ($2/5$,8) for $L{\leq}11$; ($2/3$,6) for $L{\leq }4$; ($2/3$,8) for $L{\leq}4$; ($2/7$,6) for $L{\leq }4$; ($2/7$,8) for $L{\leq}4$; ($2/9$,6) for $L{\leq}4$; ($2/9$,8) for $L{=}2$; bosonic ($2/3$,6) for $L{\leq}4$; bosonic ($2/3$,8) for $L{\leq}4$. We find ${\cal O}{=}1$ in arbitrary precision calculations for all of these. This is nontrivial, because, as shown in Table~\ref{tab: GMP_in_CFE}, the number of CFEs at a given $L$ is $N$-independent (provided $N$ is sufficiently large) and much smaller than the number of basis functions, which ranges to several thousand for the systems studied and increases exponentially with $N$. This gives us confidence that the GMP mode is generally contained entirely within the (direct-projected) CFE subspace, although we have not yet succeeded in devising a general proof.

\begin{table}[t!]
	\begin{center}
		\begin{tabular}{ |c|c|c|c|c|c|c|c|c| } 
			\hline
			$\nu$ & $N$ & & $L{=}0$ & $L{=}2$ & $L{=}3$ & $L{=}4$ & $L{=}5$ & $L{=}6$ \\ \hline \hline
			$1/3$ 				&  10  & D$_{L}$  & 319        & 1,357      & 1,769     & 2,371      & 2,752 & 3,330 \\ \hline
		            		&        & D$^{\rm CFE}_{L}$ &       1(1)         & 1(2)      & 1(3)     & 2(4)      &2(5) &3(6) \\ \hline
			  &              & ${\cal O}_{\rm JK}$ &     & 1    & 1  & 1   & 1  & 0.9999\\ \hline \hline
			$2/5$ 				&  12  & D$_{L}$   	& 418        & 1,875      & 2,511     & 3,296      & 3,885 & 4,638 \\ \hline
			&          &    D$^{\rm CFE}_{L}$ &   1(1)         & 2(3)      & 4(5)     & 6(7)      &8(9)    &10(11) \\ \hline
			&          & ${\cal O}_{\rm JK}$     &   & 0.9994    & 0.9996  & 0.9996   & 0.9998 & 0.9998  \\ \hline \hline
			$3/7$ 				&  12   & dim$_{L}$ 	& 127        &  493      & 621    & 852      & 961      & 1,182 \\ \hline
			&         & D$^{\rm CFE}_{L}$  &       1(1)         & 2(3)      & 5(6)     & 8(9)      &10(11)      &11(12) \\ \hline
			&         & ${\cal O}_{\rm JK}$       &     & 0.9981    & 0.9969  & 0.9982   & 0.9990  & 0.9994  \\ \hline 
		\end{tabular}
	\end{center}
	\caption{\label{tab: GMP_in_CFE}This table gives ${\cal O}_{\rm JK}$, the sum of the squared overlaps of the GMP mode with the single CF-exciton states obtained with the Jain-Kamilla (JK) projection, for the largest systems studied. D$_L$ (D$^{\rm CFE}_{L}$) is the number of linearly independent basis (CFE) states at $L$; the number of excitons at $Q^*$ is shown in brackets. When shown as ${\cal O}{=}1$, ${\cal O}$ is 1 to machine precision.}

\end{table}

The values of ${\cal O}_{\rm JK}$ obtained for the JK-projected CFEs are given in Table~\ref{tab: GMP_in_CFE} for $\nu{=}n/(2n{+}1)$. In all cases studied, we find that ${\cal O}_{\rm JK}{>}0.997$, demonstrating that Eq.~\eqref{conjecture1} is practically exact for the JK projection. The JK-projected CFEs for $n/(4n{\pm} 1)$ Jain states are considered below.

The relation in Eq.~\eqref{conjecture1} between the GMP mode and single CFEs is unexpected as these are motivated by qualitatively different physical descriptions. The GMP ansatz creates a coherent superposition of states containing a single electron-hole pair while a CFE involves the excitation consisting of a single CF particle and a single CF hole, each of which, in isolation, has a fractional charge of magnitude $e/(2n{\pm}1)$~\cite{Jain07}. Indeed, the dispersion of the GMP mode is generally rather different from that of the primary CFE~\cite{Kamilla96b, Kamilla96c, Scarola00, Dora24}. Moreover, because an electron added to a FQH state is described as a complex bound state of $2n{\pm} 1$ excited CFs~\cite{Jain05, Pu22, Pu23a, Gattu23}, one would a priori expect an electron-hole pair of the GMP mode to involve $2n{\pm} 1$ CFEs. Eq.~\eqref{conjecture1} is far from obvious also from energetic considerations (see SM~\cite{SM_Balram_Sreejith_Jain_Graviton}).

\textbf{\textit{CFE ansatz for $S(\vec{q},E)$.}} Substituting Eq.~\eqref{conjecture1} into  Eq.~\eqref{eq: Sqomega} suggests that the SMA of Eq.~\eqref{eq: SMA} can be replaced by a ``CFE Ansatz" for the LLL-projected DSF:
\begin{equation}\label{eq: CFA}
S(\vec{q},E)=\sum_{\alpha}S^{\rm CFE}_{\vec{q}\alpha}\delta(E-E^{\rm CFE}_{\vec{q},\alpha})
\end{equation}
with $S^{\rm CFE}_{\vec{q},\alpha}{=}|\langle \Psi_{\vec{q},
\alpha}^{\rm CFE}|\Psi_{\vec{q}}^{\rm GMP}\rangle|^2$, where $\Psi_{\vec{q},\alpha}^{\rm CFE}$ and $E^{\rm CFE}_{\vec{q},\alpha}$ are the CFE eigenstates and eigenenergies obtained by diagonalizing the Coulomb interaction in the CFE basis. The GMP mode thus splits into a ladder of discrete CFE modes, as depicted schematically in Fig.~\ref{fig: schematic}(h, i).

\textbf{\textit{Graviton.}} 
The neutral mode at $L{=}2$ is excited in inelastic light scattering experiments with circular polarization~\cite{Liang24}. The GMP and CFE at $L{=}2$ are referred to as GMP and CF gravitons. 

\begin{figure*}[t]
	\centering    
	\includegraphics[width=0.32\linewidth]{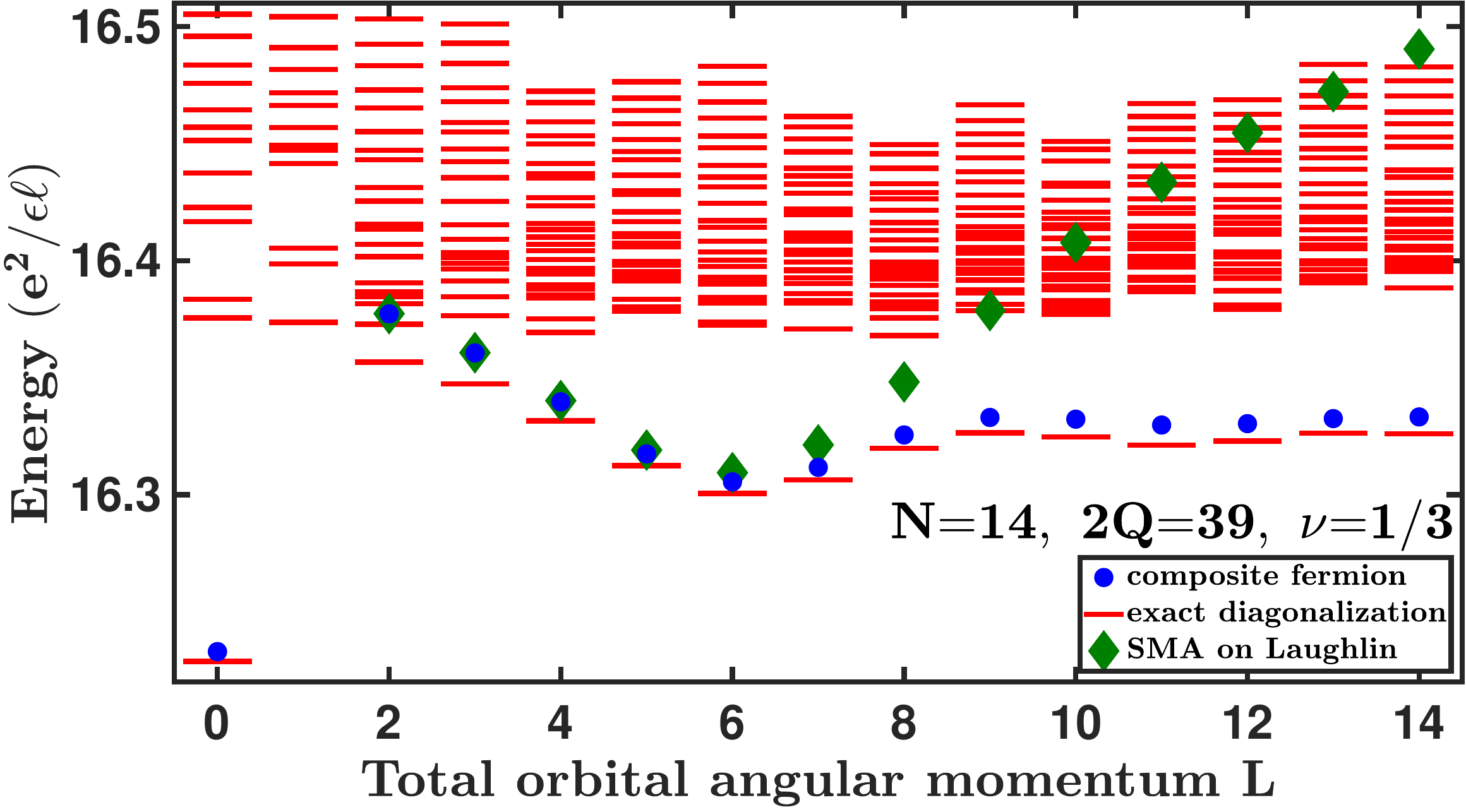}
	\includegraphics[width=0.32\linewidth]{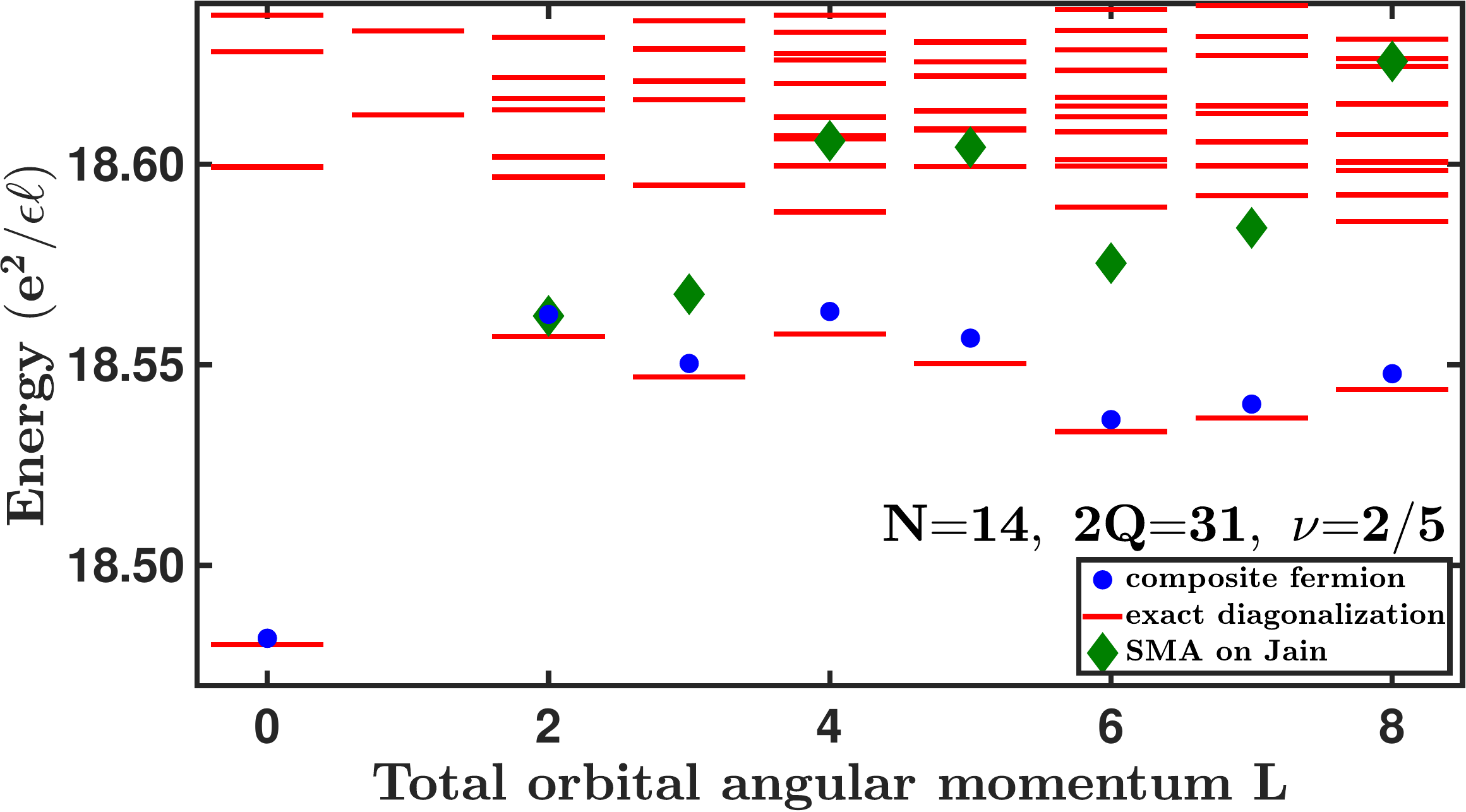}
	\includegraphics[width=0.32\linewidth]{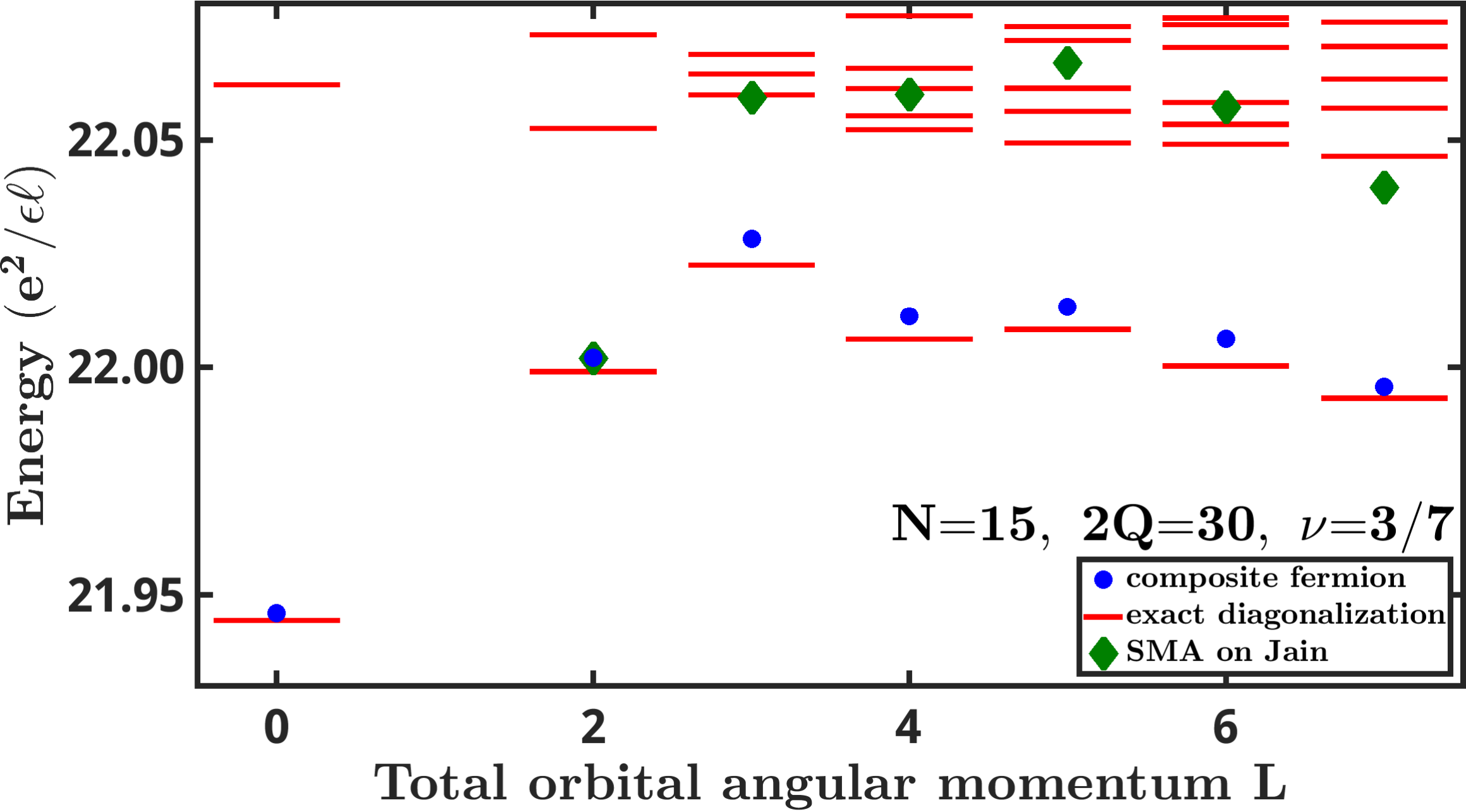} \\
        \includegraphics[width=0.32\linewidth]{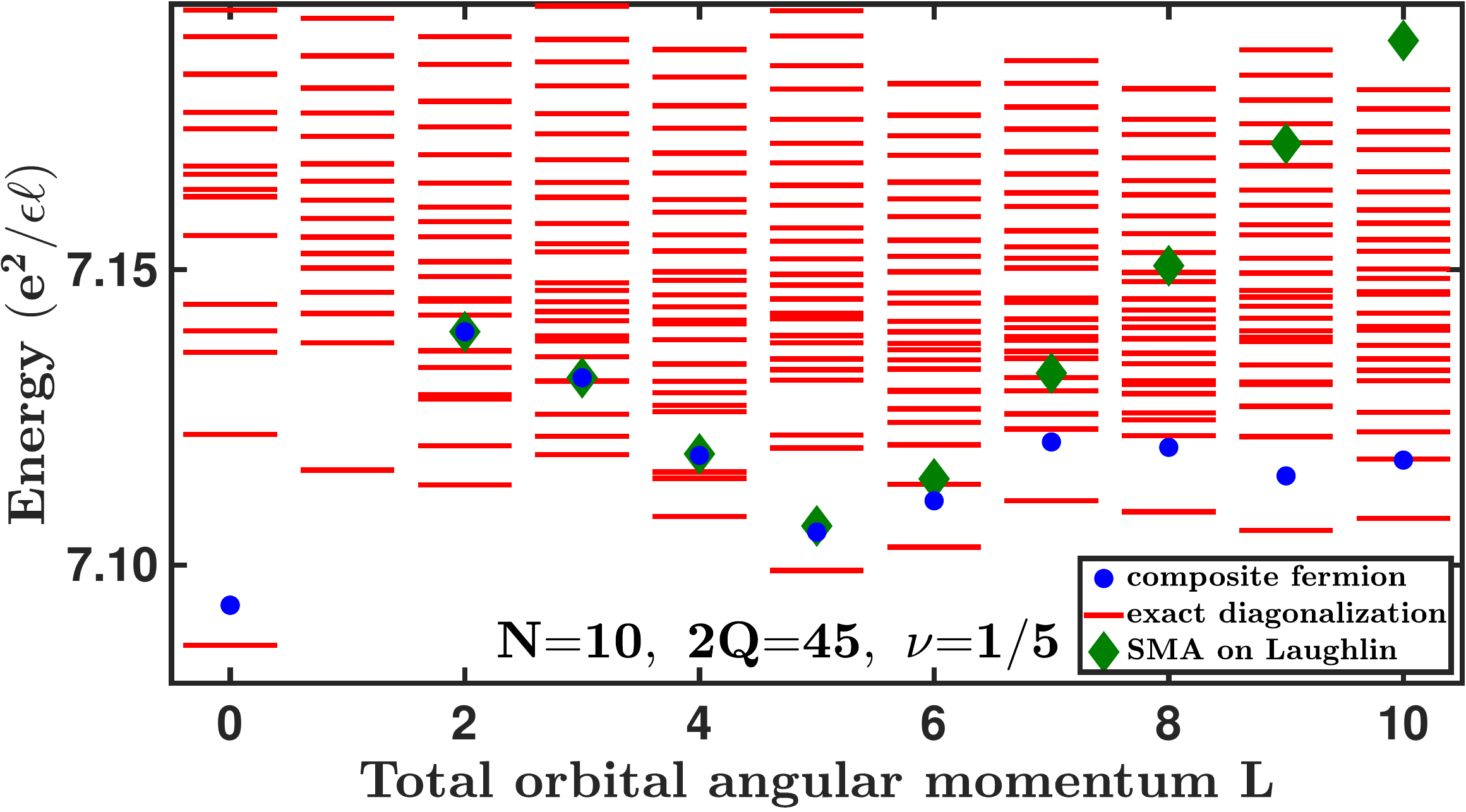}
	\includegraphics[width=0.32\linewidth]{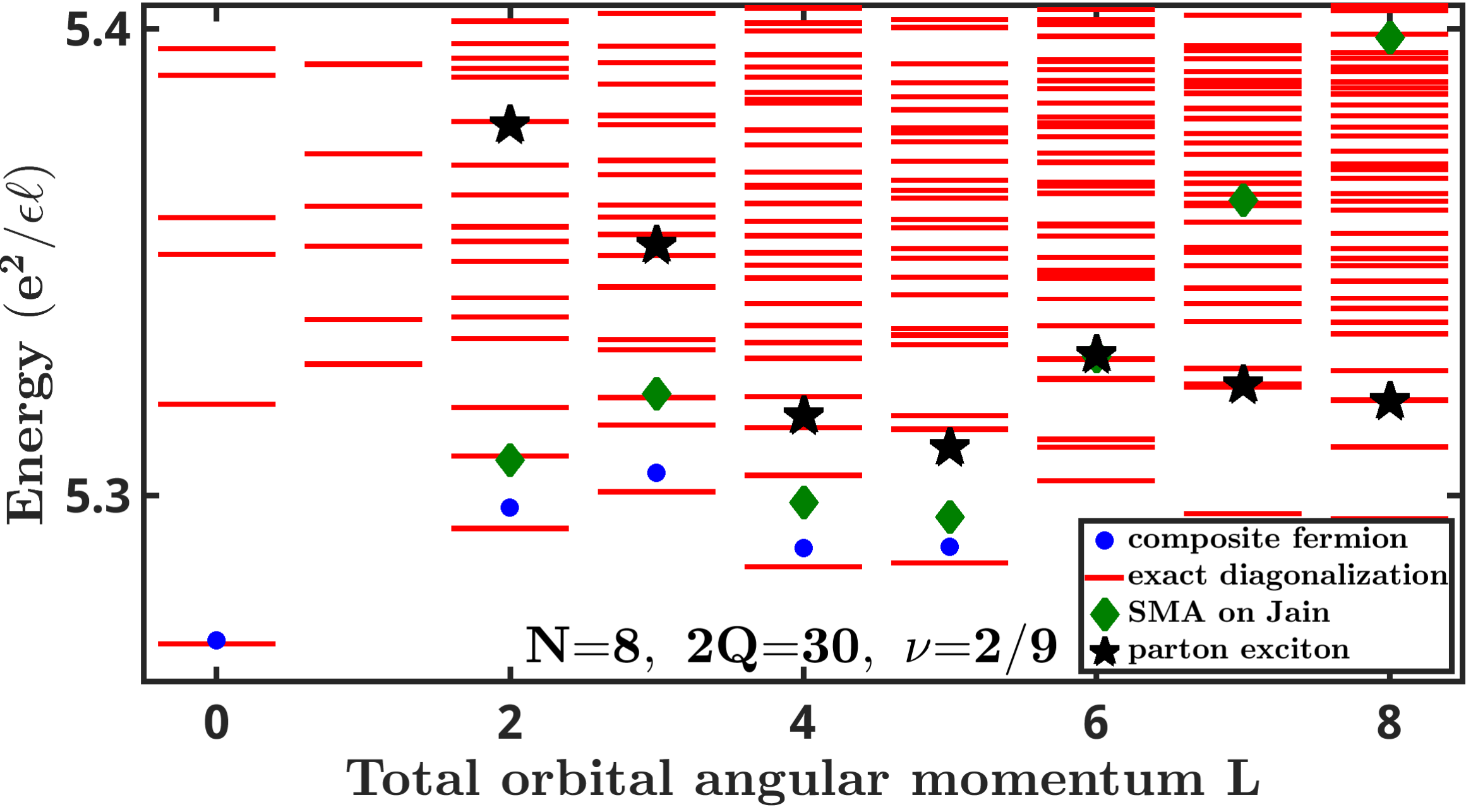}
	\includegraphics[width=0.32\linewidth]{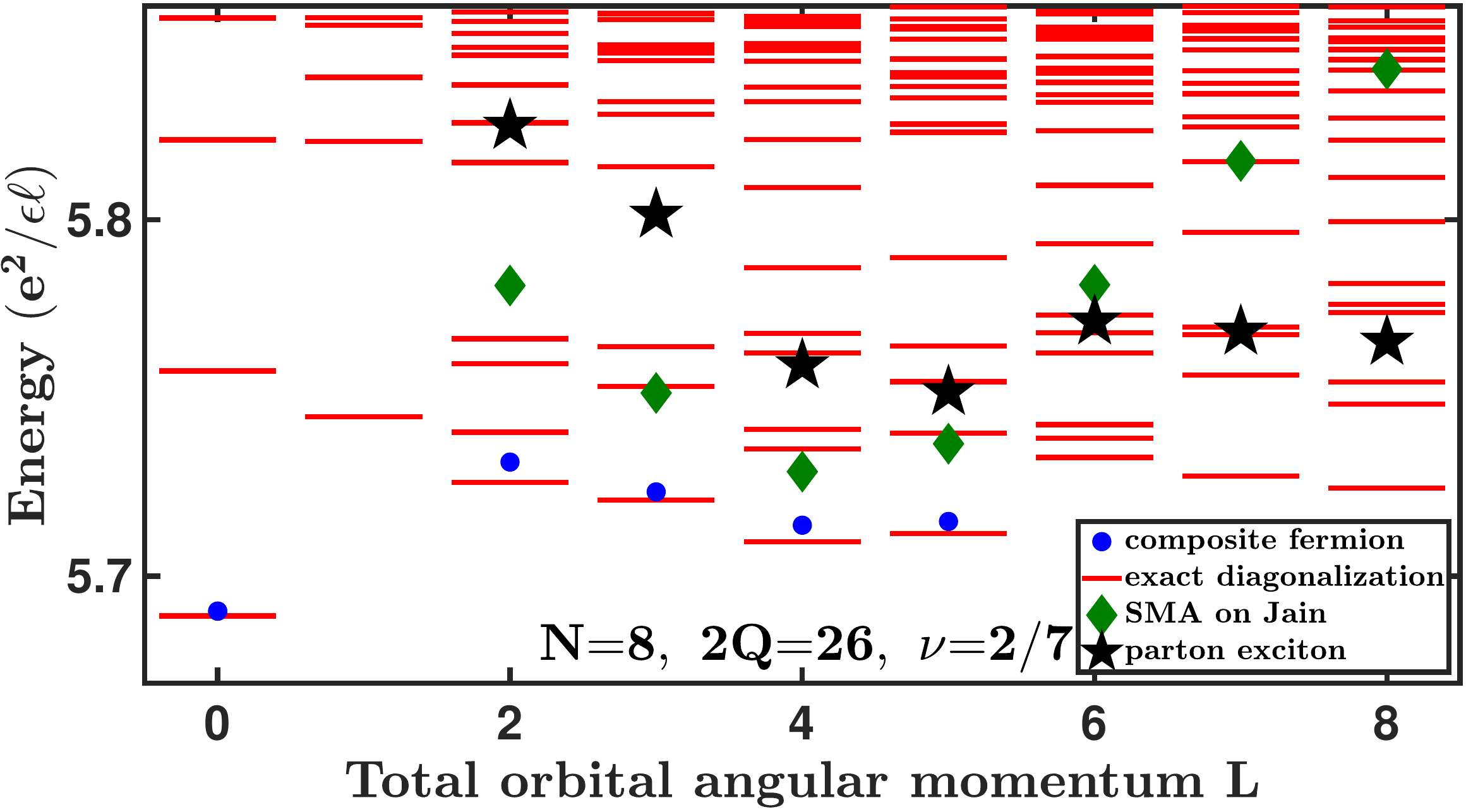}
	\caption{\label{fig: GMP_Laughlin_Jain_CFE_partonE_LLL_Coulomb} The energies of the GMP mode (green diamonds), the JK-projected primary CF exciton (blue dots) for $\nu{=}1/3,~2/5,~3/7,~1/5,~2/9$ and $2/7$, and the JK-projected primary parton exciton (black pentagrams) for $\nu{=}2/7, 2/9$. Red dashes show the exact Coulomb spectrum. The spherical geometry is used, and the GMP energy is computed using the Laughlin (Jain) ground state for $1/3,~1/5$ ($2/5,~3/7,~2/9,~2/7$).
	}
\end{figure*}

We find that for $\nu{=}n/(2n{\pm}1)$ and $\nu{=}1/5$, the GMP graviton is very well approximated by the primary CF graviton, as seen in Fig.~\ref{fig: GMP_Laughlin_Jain_CFE_partonE_LLL_Coulomb}. This figure also shows the Coulomb energies of the primary CFE and the GMP mode obtained using the Laughlin/Jain ground states; these are identical for $L{=}2,3$ for 1/3 and 1/5~\cite{Kamilla96c, Yang12b, Pu24} and very close for $L{=}2$ for 2/5 and 3/7. For comparison, we also show spectra obtained from exact diagonalization (ED) of the Coulomb interaction taken from Refs.~\cite{Balram13, Jain14}. The energy-resolved spectral function of the graviton essentially has a single peak at these fillings. 

\begin{figure}[t]
	\centering    
	\includegraphics[width=\linewidth]{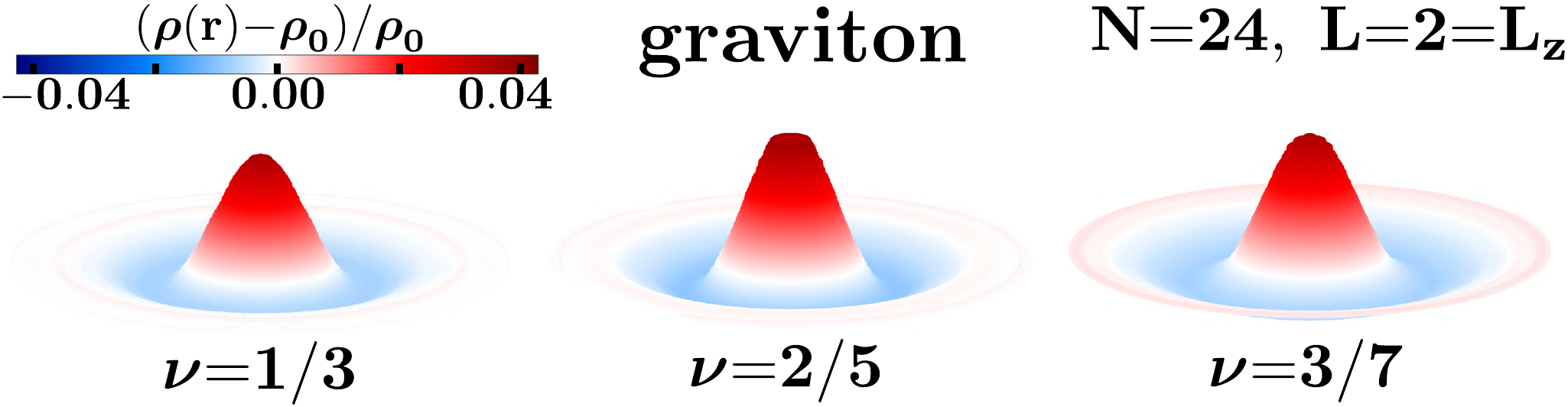}
	\caption{\label{fig: graviton_density}Plot of $(\rho(r){-}\rho_0)/\rho_0$ for the JK-projected CF graviton with $L{=}2{=}L_{z}$ at $\nu{=}1/3$, $2/5$ and $3/7$, where $\rho(r)$ and $\rho_0$ are the densities of the graviton and the ground state, and $r$ is the distance from the center.  
}
	
\end{figure}

The identification with the $L{=}2$ CFE allows a study of the graviton in large systems (which is not possible for the GMP mode). Figure~\ref{fig: graviton_density} shows the density $\rho(r)$ of the CF graviton at $\nu{=}1/3$, $2/5$ and $3/7$ for $N{=}24$. The graviton size increases with $n$ along $n/(2n{+}1)$, as expected, given that the size of the constituent CF particle and CF hole is of order $(2n{+}1)\;\ell$~\cite{Jain07, Balram13b, Gattu23}.

\begin{figure}[t]
	\centering    
	\includegraphics[width=\columnwidth]{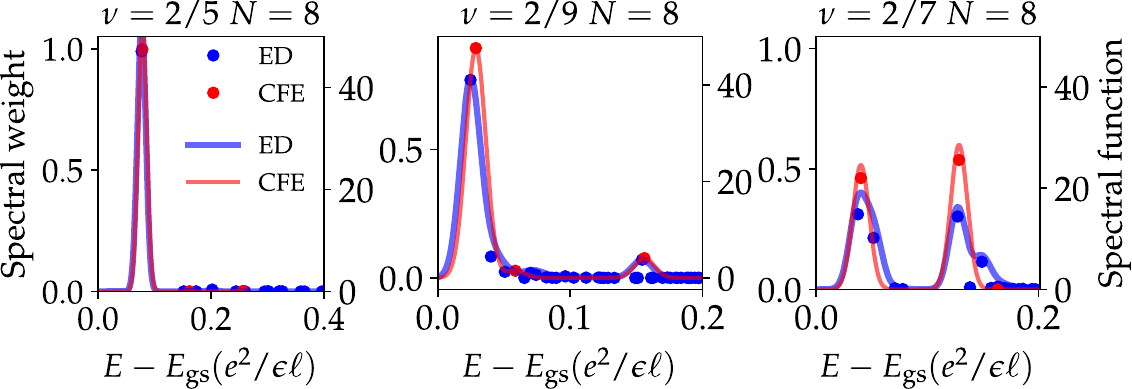}
	\caption{\label{fig: SWdirectProjection} Comparison between the LLL-projected DSFs for the $L{=}2$ ``graviton" at $\nu{=}2/5$, $2/7$ and $2/9$ obtained from exact diagonalization [Eq.~\eqref{eq: Sqomega}] (blue lines) and the CFE Ansatz [Eq.~\eqref{eq: CFA}] (red lines), with the delta functions broadened into Gaussians of width 0.0075${e^2}/{\ell}$ (SM). Blue (red) dots show spectral weights $|\langle\Psi^{\rm ED}_\alpha|\Psi_{L{=}2}^{\rm GMP}\rangle|^2$ ($|\langle\Psi^{\rm CFE}_j|\Psi_{L=2}^{\rm GMP}\rangle|^2$) and energies are measured relative to $E_{gs}$, where $E_{\rm gs}$ is the ground state energy, $\Psi^{\rm ED}_\alpha$ are the exact Coulomb eigenstates, $\Psi^{\rm CFE}_j$ are the Coulomb eigenstates obtained within the three-dimensional direct-projected CFE basis. The GMP state is obtained using the CF ground state.}
\end{figure}
 
For the Jain states at $\nu{=}n/(4n{\pm} 1)$ with $n{>}1$ the energy of the GMP graviton differs significantly from that of the primary CF graviton, as seen in Fig.~\ref{fig: GMP_Laughlin_Jain_CFE_partonE_LLL_Coulomb}. Fig.~\ref{fig: SWdirectProjection} shows the spectral weight obtained by diagonalizing the Coulomb interaction within the direct-projected CF graviton space (which has three linearly independent modes). It is in excellent agreement with that obtained by ED, supporting the validity of Eq.~\eqref{eq: CFA}. The presence of two prominent peaks indicates that the GMP graviton splits into two CF gravitons for a range of broadening.

Remarkably, Nguyen and Son~\cite{Nguyen21} predicted an additional high-energy graviton at $n/(2pn{\pm}1)$ for $2p{\geq} 4, n{\geq}2$, motivated by the observation that a Dirac CF field theoretical description of $\nu{=}n/(4n{\pm} 1)$ states by Goldman and Fradkin~\cite{Goldman18} and Wang~\cite{Wang19} does not satisfy a bound derived by Haldane~\cite{Haldane11a} which says that the coefficient of the $q^4$ term in the projected static structure factor is lower-bounded by $({\cal S}{-}1)/8$, where ${\cal S}$ is the Wen-Zee shift~\cite{Wen92}. 
Balram \emph{et. al.}~\cite{Balram21d} proposed to identify the second, higher energy graviton as a ``parton graviton." In the parton construction of the FQH states ~\cite{Jain89b} one divides each electron into an odd number of unphysical fractionally charged fermions called partons [see Fig.~\ref{fig: schematic}(j)], places each species of partons in an integer quantum Hall state, and then combines the partons back into physical electrons. In the Jain $n/(2pn{\pm} 1)$ states, one species of partons forms the $\nu{=}{\pm} n$ state and remaining $2p$ partons the $\nu{=}1$ state. Balram \emph{et. al.}~\cite{Balram21d} defined the parton exciton as the exciton in $\Phi_1$ i.e., 
\begin{equation}
	\label{eq: partonE}
	\Psi^{{\rm parton-ex}}_{\vec{q},0\rightarrow K}={\cal P}_{\rm LLL}[\rho_{_{\vec{q}}}^{0\rightarrow K} \Phi_1] \Phi_1^{3} \Phi_{\pm n}
\end{equation}
[see Fig.~\ref{fig: schematic}(c)]. For the Laughlin $1/(2p{+}1)$ states~\cite{Laughlin83}, the parton exciton is the same as the primary CFE. 

Surprisingly, Eq.~\eqref{conjecture1} is not a good approximation for the JK projected CFEs for the Jain $\nu{=}n/(4n{\pm} 1)$ states with $n{>}1$. For example, for $\nu{=}2/7$ ($\nu{=}2/9$) we find ${\cal O}_{\rm JK}{=}0.5076$ (${\cal O}_{\rm JK}{=}0.9191$) for $N{=}8$ at $L{=}2$. [To project, we write $\Psi_{n/(4n{\pm} 1)}{=}\Phi_1^2 \Psi_{n/(2n{\pm} 1)}$; 2/7 has a single CF graviton because 2/3 does (the hole conjugate of the single graviton at 1/3), and 2/9 inherits 2 CF gravitons from 2/5.] It turns out that the GMP graviton lies, to an excellent approximation, within the 2-dimensional subspace defined by the primary CF and parton gravitons. For the $N{=}8$ particle $\nu{=}2/7$ ($\nu{=}2/9$) state, the GMP graviton has ${\cal O}_{\rm JK}{=}0.9995$ (${\cal O}_{\rm JK}{=}0.9986$) with this subspace. Using configurations of left and right circularly polarized light, the graviton chirality is measured to be negative for the $n/(2n{+}1)$ states while it is positive for the $n/(2n{-}1)$ states~\cite{Liang24}. Within the parton description, the chirality of the graviton~\cite{Liou19} is negative (positive) if the effective magnetic field seen by the parton hosting the exciton is parallel (antiparallel) to the applied field~\cite{Balram21d}.

\textbf{\textit{Discussion.}} 
The CFEs should in principle be observable by light scattering. We expect that for the $\alpha$ CFE mode, $S^{\rm CFE}_{\vec{q},\alpha}$ is maximized at wave vectors where its energy intersects the GMP mode. Thus, in place of a single GMP mode, we expect many discrete modes, highlighted in the schematic Fig.~\ref{fig: schematic}. Evidence of multiple modes has been seen in numerical studies~\cite{He96, Majumder09}. For a disorder-free system Raman scattering probes only small  $q\ell$ modes, but an external one-dimensional periodic potential, e.g., that produced by a surface grating or by piezoelectric coupling to a surface phonon, can help select modes with finite wave vectors~\cite{Kukushkin07, Kukushkin09} and should reveal the discrete CF modes. Larger $q\ell$ can also be accessed in tilted field Raman experiments at low electron densities. While there is only one graviton for $\nu{=}1/3$, its splitting into two at small but non-zero $q$ was observed in Ref.~\cite{Hirjibehedin05} and explained in Ref.~\cite{Majumder09} in terms of CFEs. 

Disorder can also activate finite wave vector modes and produce Raman peaks at the minima and maxima of the CFE dispersion where the density of states has a peak~\cite{Pinczuk88, Kang00, Kang01, Groshaus08}.  Certain peaks in Raman spectra at 1/3 and 2/5 have been identified with the minima in the dispersion (called roton minima)~\cite{Kang00, Kang01}. High-energy Raman peaks have been observed at $1/3$~\cite{Hirjibehedin05, Rhone11} and their energies are consistent with CFE modes involving CF excitations across multiple $\Lambda$Ls~\cite{Majumder09, Rhone11}. However, an experimental determination of the mode energies as a function of $\vec{q}$ will be necessary for a decisive confirmation.

In summary, we have shown that the GMP density wave mode splits into a ladder of CFE modes, as shown in Fig.~\ref{fig: schematic}(h,i), and for $\nu{=}n/(4n{\pm }1)$ with $n{>}1$ the GMP graviton splits into two CF gravitons, also understandable as primary CF and parton gravitons. An experimental confirmation of these modes should provide further insight into the structure of the FQH states. 

\textbf{\textit{Acknowledgments.}} 
We acknowledge useful discussions with Abhishek Anand, Rakesh Dora, Lingjie Du, Mytraya Gattu, Zlatko Papi\'c, Songyang Pu, and Arkadius\'z W\'ojs. This work was undertaken on the Nandadevi supercomputer maintained and supported by the Institute of Mathematical Science's High-Performance Computing Center, India. Some of the numerical calculations were performed using the DiagHam libraries~\cite{DiagHam}. ACB thanks the Science and Engineering Research Board (SERB) of the Department of Science and Technology (DST) for funding support via the Mathematical Research Impact Centric Support (MATRICS) Grant No. MTR/2023/000002. JKJ was supported in part by the U. S. Department of Energy, Office of Basic Energy Sciences, under Grant No. DE-SC-0005042. We thank the National Supercomputing Mission (NSM) for providing computing resources of ‘PARAM Brahma’ at IISER Pune, which is implemented by C-DAC and supported by the Ministry of Electronics and Information Technology (MeitY) and the Department of Science and Technology (DST), Government of India. 

\bibliography{biblio_fqhe}

\newpage 
\cleardoublepage

\onecolumngrid
\begin{center}
\textbf{\large Supplemental Material for ``Splitting of Girvin-MacDonald-Platzman density wave and the nature of chiral gravitons in fractional quantum Hall effect"}\\[5pt]

\begin{center}
 {\small Ajit C. Balram$^{1,2}$, G. J. Sreejith$^{3}$ and J. K. Jain$^{4}$ }  
\end{center}

\begin{center}
{\sl \footnotesize
$^{1}$Institute of Mathematical Sciences, CIT Campus, Chennai 600113, India

$^{2}$Homi Bhabha National Institute, Training School Complex, Anushaktinagar, Mumbai 400094, India
    
$^{3}$Indian Institute of Science Education and Research, Pune, India 411008

$^{4}$Department of Physics, 104 Davey Lab, Pennsylvania State University, University Park, Pennsylvania 16802, USA
}
\end{center}

\begin{quote}
{\small This supplemental material contains certain technical details that are not included in the main text. In Sec.~\ref{sec: background}  we provide a primer on the Girvin-MacDonald-Platzmann (GMP) and composite-fermion exciton (CFE) modes. Then, in Sec.~\ref{sec: direct_projection_results} we present the method of and results obtained from direct projection. Energies of the GMP density wave mode obtained from acting the projected density operator on the exact LLL Coulomb ground state are presented in Sec.~\ref{sec: GMP_Coulomb}. (In Fig. 3 of the main text we presented GMP energies obtained from acting the projected density operator on the model Laughlin/Jain ground states.). Sec.~\ref{sec: disorder_broadening_DSF} displays the effect of disorder-broadening or the resolution of the measurement on the spectral function. The last section, Sec.~\ref{sec: DSF_L_geq2_2_5}, presents the dynamical structure factor for total orbital angular momentum $L{\geq}2$ at $\nu{=}2/5$. 
}
\end{quote}
\end{center}

\vspace*{0.4cm}

\setcounter{equation}{0}
\setcounter{figure}{0}
\setcounter{table}{0}
\setcounter{page}{1}
\setcounter{section}{0}
\makeatletter
\renewcommand{\theequation}{S\arabic{equation}}
\renewcommand{\thefigure}{S\arabic{figure}}
\renewcommand{\thesection}{S\Roman{section}}
\renewcommand{\thepage}{\arabic{page}}
\renewcommand{\thetable}{S\arabic{table}}

\section{Background}
\label{sec: background}
\paragraph{Density operator}
The density operator in the first quantized notation is given by 
\begin{equation}
\rho (\theta,\phi) = \sum_{i=0}^N \delta (\cos \theta-\cos \theta_i) \delta (\phi - \phi_i) =  \sum_{l=0}^\infty \sum_{m=-l}^l Y^*_{0lm} (\theta,\phi) \sum_{i=1}^N Y_{0lm} (\theta_i \phi_i)
\end{equation}
where $Y_{0lm}$, a special case of the ``monopole harmonics" $Y_{qlm}$, are the same as the usual spherical harmonics $Y_{l,m}$.
Identifying this as a mode expansion in spherical harmonics we define the $l,m$ mode of the density operator as 
\begin{equation}
\rho_{L=l,L_z=m} = \sum_{i=1}^N Y_{0lm} (\theta_i \phi_i)
\end{equation}
It is easy to check that $\rho_{lm}$ adds an angular momentum  $L{=}l,~L_z{=}m$ to the state it acts on. In particular, it produces a many-body state with $L{=}l$ and $L_z{=}m$ when it acts on the ground state (which is a rotation-invariant state $L{=}0$).  

Since $Y_{000}{\sim} 1$, action of $\rho_{0,L{=}0,m{=}0}$ on the ground state returns the ground state. In general, the spectral functions being calculated here are rotation invariant, therefore it is sufficient for our purposes to consider the lowest weight state ($L_z{=}{-}L$) of each angular momentum multiplet. When the spherical harmonics are expressed in terms of the spinor coordinates $u_{i}{=}\cos(\theta_{i}/2)e^{i\phi/2},~v_{i}{=}\sin(\theta_{i}/2)e^{{-}i\phi_{i}/2}$, the highest weight density operators have the form $Y_{Q, L, L_{z}}(u_{i}, v_{i}){\propto}v^{L}_{i}\bar{u}^{L}_{i}$. The corresponding LLL projected density operator is obtained by replacing $\bar{u}$ with $\partial/\partial_{u}$ and normal-ordering the derivatives, i.e., moving all the derivatives to the left. With the explicit form of the projected density operator, one can then show that the $L{=}1$ GMP mode is annihilated upon projection to the LLL~\cite{Dora24, He94}. Thus the GMP mode starts from $L{=}2$ (graviton) and extends up to $L^{\rm GMP}_{\rm max}{=}2Q{\sim}N/\nu$ for large $N$ because the GMP operator creates a coherent superposition of states containing an {\it electron} and a {\it hole} in the LLL each of which carries a single-particle angular momentum of $l{=}Q$.

For the CFE state defined in Eq. (3) of the main text, $\Phi_{n}$, the wave function of $n$ filled LLs is constructed at an effective flux of $2Q^{*}{=}2Q{-}2(N{-}1){=}N/n{-}n$. There are many CFE modes corresponding to excitations in $\Phi_n$ across different LLs. The lowest energy CFE called the ``primary CFE", corresponds to the $n{-}1{\rightarrow} n$ excitation. The angular momenta of the constituent CF particle and CF hole are $l^{\rm CFP}{=}Q^{*}{+}n$ and $l^{\rm CFH}{=}Q^{*}{+}n{-}1$. The primary CFE extends from $L^{\rm CFE}_{\rm min}{=}2$ (the $L{=}1$ CFE is also annihilated upon projection into the LLL~\cite{Dev92}) to $L^{\rm CFE}_{\rm max}{=}l^{\rm CFH}{+}l^{\rm CFP}{=}2Q^{*}{+}2n{-}1{=}(N/n{+}n{-}1){\sim}N/n{=}N/\nu^{*}$ for large $N$, where $\nu^{*}$ is the effective filling of CFs~\cite{Balram16d}. In general, the $m{\rightarrow} m{+}K$ CFE mode starts from $L{=}K$ (for certain CFEs the $L{=}K$ excitation is annihilated upon projection to the LLL~\cite{Dev92, Balram13} in which case the mode starts from $L{=}K{+}1$) and extends up to $2Q^{*}{+}2m{+}K$ since the CF particle and CF hole angular momenta are $l^{\rm CFP}{=}Q^{*}{+}m{+}K$ and $l^{\rm CFH}{=}Q^{*}{+}m$. As an example, at $L{=}2$, there are at most two CFEs for $\nu{=}1/3$ [$0{\rightarrow}1$ and $0{\rightarrow} 2$ excitons (Note that the $0{\rightarrow} 3$ exciton has $L{\geq}3$.)], and at most three for $\nu{=}2/5$ ($1{\rightarrow} 2$, $1{\rightarrow} 3$ and $0{\rightarrow}2$). An explicit construction of the wave functions is required to get the final counting of states at a given $L$ since certain CFEs can be annihilated upon projection to the LLL. The states at $\nu{=}(n{+}1)/(2n{+}1)$ are related by particle-hole symmetry to the states at $\nu{=}n/(2n{+}1)$~\cite{Girvin84}. Thus, the spectra at the two fillings, particularly their gravitons, can be related to each other. For example, the 2/3 state has only one graviton since its hole conjugate at 1/3 supports only a single graviton. Furthermore, the chirality of the graviton at $(n{+}1)/(2n{+}1)$ is opposite to the chirality of the graviton at $n/(2n{+}1)$~\cite{Liou19, Balram21d, Nguyen22}. Since the secondary Jain states at $n/(4n{\pm}1)$ can be viewed as a product of the bosonic $1/2$ Laughlin and $n/(2n{\pm }1)$ Jain states, the $n/(4n{\pm}1)$ state inherits (by inherit, here we mean if a LLL FQH ground state wave function $\Psi$ can be written as a product of two LLL FQH ground state wave functions $\Psi^{\alpha}$ and $\Psi^{\beta}$, and $\Psi^{\alpha}$ and $\Psi^{\beta}$ have some particle-hole like modes, we generally expect $\Psi$ to inherit them) the gravitons of the $n/(2n{\pm }1)$ Jain state (additional graviton inherited from the bosonic $1/2$ Laughlin state can also be present).

\vspace{0.5cm}
Before proceeding to the results of our calculations, we ask whether the relation given in Eq.~(4) of the main text may be anticipated from energetic considerations. Without explicit calculation, the extent to which the GMP mode and the CF excitons (CFEs) describe the low-energy excitations is unknown. Calculations show that the GMP mode is close to the lowest energy mode of the $n/(2n{\pm} 1)$ states at small wave vectors but not at large wave vectors, while for the $n/(4n{\pm} 1)$ states it is not a low energy mode at any wave vector (except for the $1/5$ state). The lowest energy CFE has been found to provide an excellent description of the lowest energy mode for all of these fractions at all wave vectors. The higher energy excitations are described by multiple CFEs; for example, excitations with 3 CF cyclotron energy can be produced by creating a single CFE, two CFEs, or 3 CFEs. From this perspective, it is not obvious, a priori, that the GMP mode can be expressed entirely in terms of \emph{single} CFE basis states.

\section{Direct projection into LLL: Method and results}
\label{sec: direct_projection_results}
\paragraph*{Method:} The method of ``direct projection" amounts to expanding the wave function in the Fock basis and retaining only the part that resides in the LLL~\cite{Dev92}. We summarize the combinatorial method used to evaluate the direct-projected wave functions. Consider the product wave function $\phi_1\phi_2$ where the two factors are Slater determinants containing $N$ particles occupying monopole harmonic single particle orbitals ($Y_{Q,l,l_z}$) at fluxes $Q_1$ and $Q_2$ respectively. The product expands to a linear combination of  $N! {\times} N!$ terms, each of which is a product of the form $\prod_{i{=}1}^N f_i(u_i,v_i)$. The single particle functions $f_i$, in turn, are products of monopole harmonics $Y_{Q_1,l,l_z}(u_i,v_i)Y_{Q_2,l',l'_z}(u_i,v_i)$, which are expanded as a linear combination of monopole harmonics in a flux of $Q_1{+}Q_2$, where the expansion coefficients can be evaluated in terms of Clebsch-Gordan coefficients (see Sec.~3.11.4 in Ref.~\cite{Jain07}). Putting these together, while keeping track of the fermionic statistics of each factor, we expand $\phi_1\phi_2$ as a linear combination of permanents of $N$ particles occupying monopole harmonic single particle orbitals in a flux of $Q_1{+}Q_2$. This strategy is recursively applied to express the product of $k$ Slater determinants $\prod_{i{=}1}^k \phi_i$ as a linear combination of Slater determinants (for $k$ odd) or permanents (for $k$ even) of particles in flux $\sum_{i{=}1}^k Q_i$. To obtain the LLL projection of the product, in the final step, we keep only the terms in the expansion that live entirely within the LLL.

Note that the direct projection method allows explicit construction of both the CF excitons and parton excitons, and with minor tweaks, also of the GMP modes. However, due to the rapid growth of the number of terms in the expansion at each step, this method can be implemented only for small systems, typically with 10 or fewer particles. 

The JK projection, on the other hand, can be implemented for hundreds of particles, which has enabled detailed quantitative comparisons between the predictions of the CF theory and experimental observations. This method has been described in detail elsewhere.

We note that both methods produce LLL wave functions, and thus there is no a priori preference between them. Past calculations have shown that the two methods produce wave functions that are extremely close but not identical. For example, the squared overlaps between the JK and direct projected Jain states at $\nu{=}2/5$ and $\nu{=}3/7$ for $N{=}12$ electrons are 0.9997 and 0.9993.

\paragraph*{Results:} 
In all cases that we studied, it was found that the GMP mode, parton excitons, and CFEs constructed using direct projection at $L{=}1$ vanish identically.  At higher $L$, we have studied the following ($\nu$, $N$) systems with direct projection: ($1/3$,6) for $L{=}2,3,4, 5, 6, 7$; ($1/3$,8) for $L{=}2,3,4, 5, 6, 7$; ($2/5$,6) for $L{=}2,3,4$; ($2/5$,8) for $L{\leq}11$; ($2/3$,6) for $L{=}2,3,4$; ($2/3$,8) for $L{=}2,3,4$; ($2/7$,6) for $L{=}2,3,4$; ($2/7$,8) for $L{=}2,3,4$; ($2/9$,6) for $L{=}2,3,4$; ($2/9$,8) for $L{=}2$; bosonic ($2/3$,6) for $L{=}2,3,4$; bosonic ($2/3$,8) for $L{=}2,3,4$. For all of these, we find ${\cal O}{=}1$ in arbitrary precision calculations, i.e. the GMP mode is exactly contained in the CFE space.

Not all CFE states are linearly independent when constructed using JK projection~\cite{Balram13} or the direct projection, but the dimension of the direct projected CFEs is found to be generally larger. 

One may ask what is the relation between the CF excitons and parton excitons. Is the latter contained within the former? For the $\nu{=}2/5$ state for all angular momenta that we considered (see previous paragraph), it was found that the direct projected parton excitons are contained within the direct projected CFE space. For the $n/(4n{\pm} 1)$ states, this is no longer true. At $\nu{=}2/7$ with $N{=}6$ and at $\nu{=}2/9$ with $N{=}6$, it was found that all direct projected parton excitons at $L{=}2$ are contained in the direct projected CFE space, but that does not remain true at larger $L$ where certain parton excitons are linearly independent of the CFEs. We note that while parton excitons cannot be constructed naturally within the JK projection for 2/5, whereas they can for 2/7 and 2/9. 

\section{GMP ansatz using the exact LLL Coulomb ground state}
\label{sec: GMP_Coulomb}

\begin{figure*}[t]
	\centering    
	\includegraphics[width=0.32\linewidth]{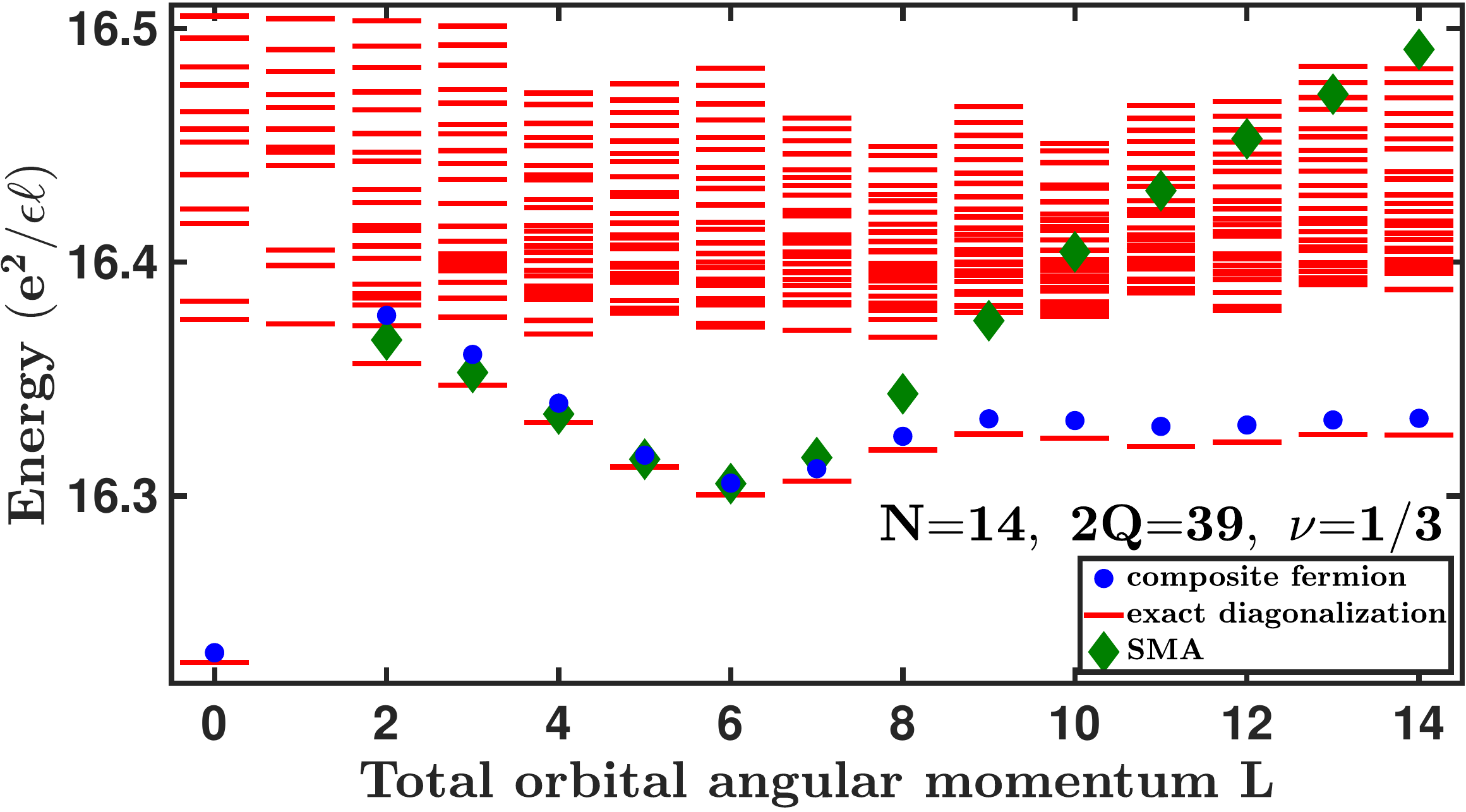}
	\includegraphics[width=0.32\linewidth]{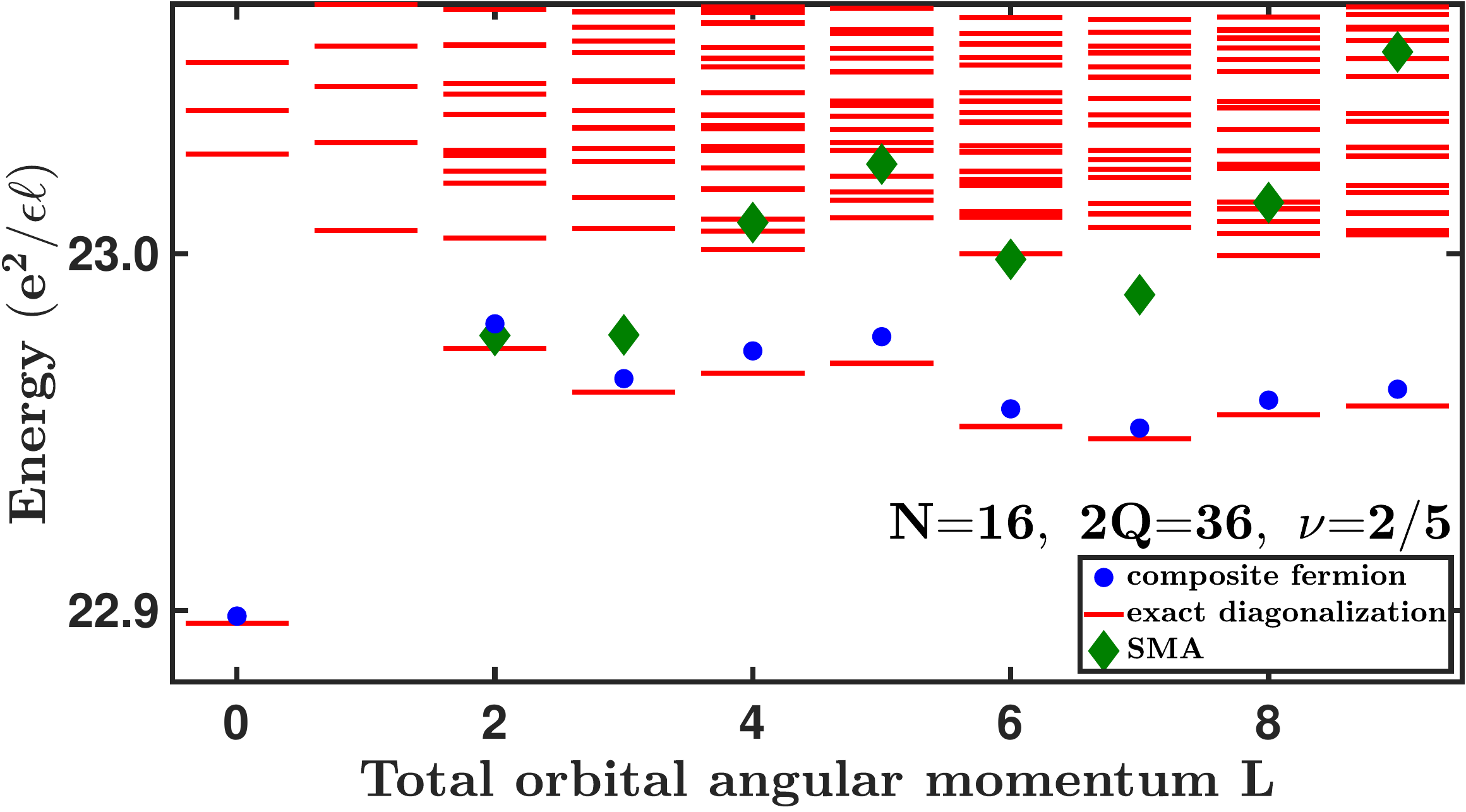}
	\includegraphics[width=0.32\linewidth]{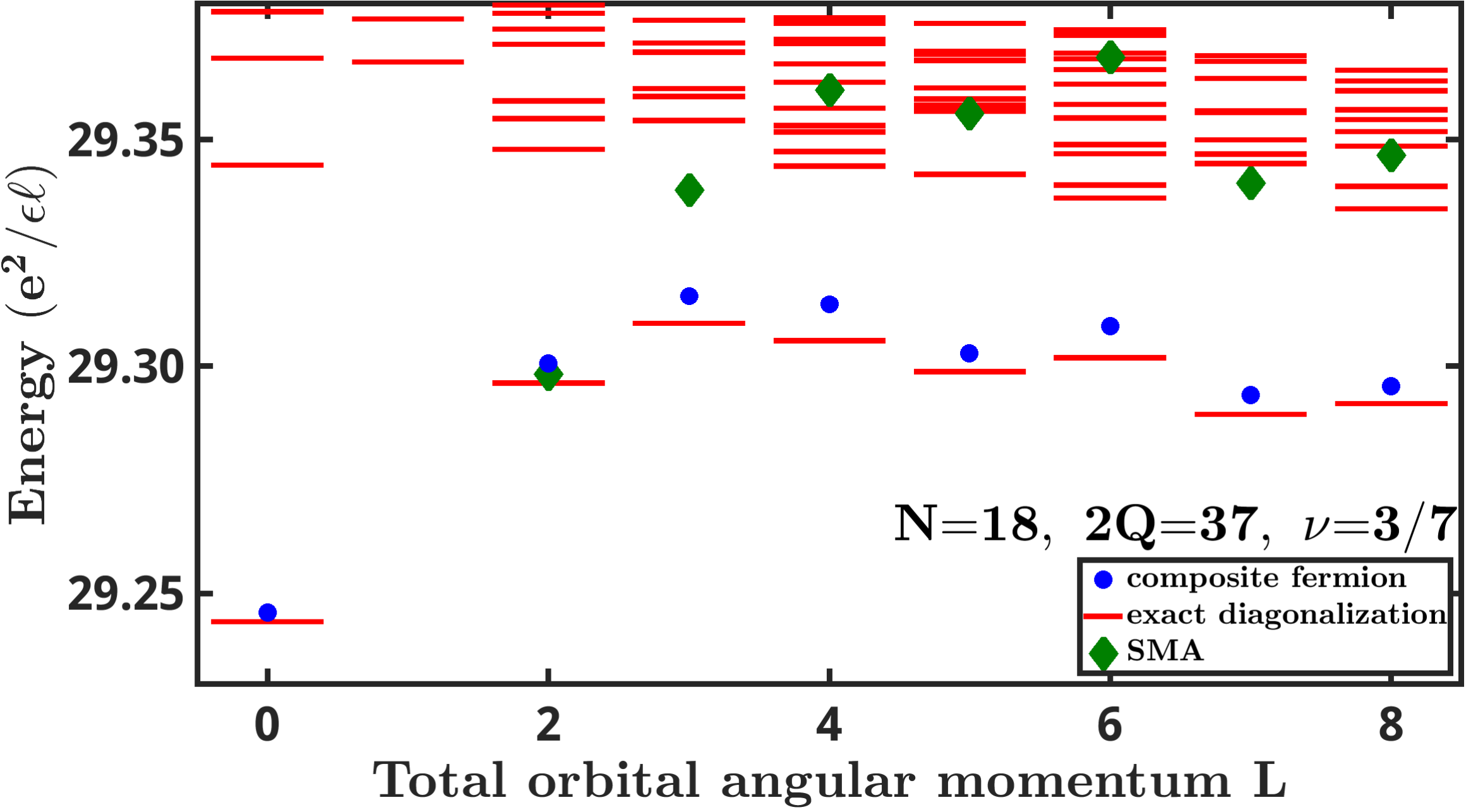} \\
        \includegraphics[width=0.32\linewidth]{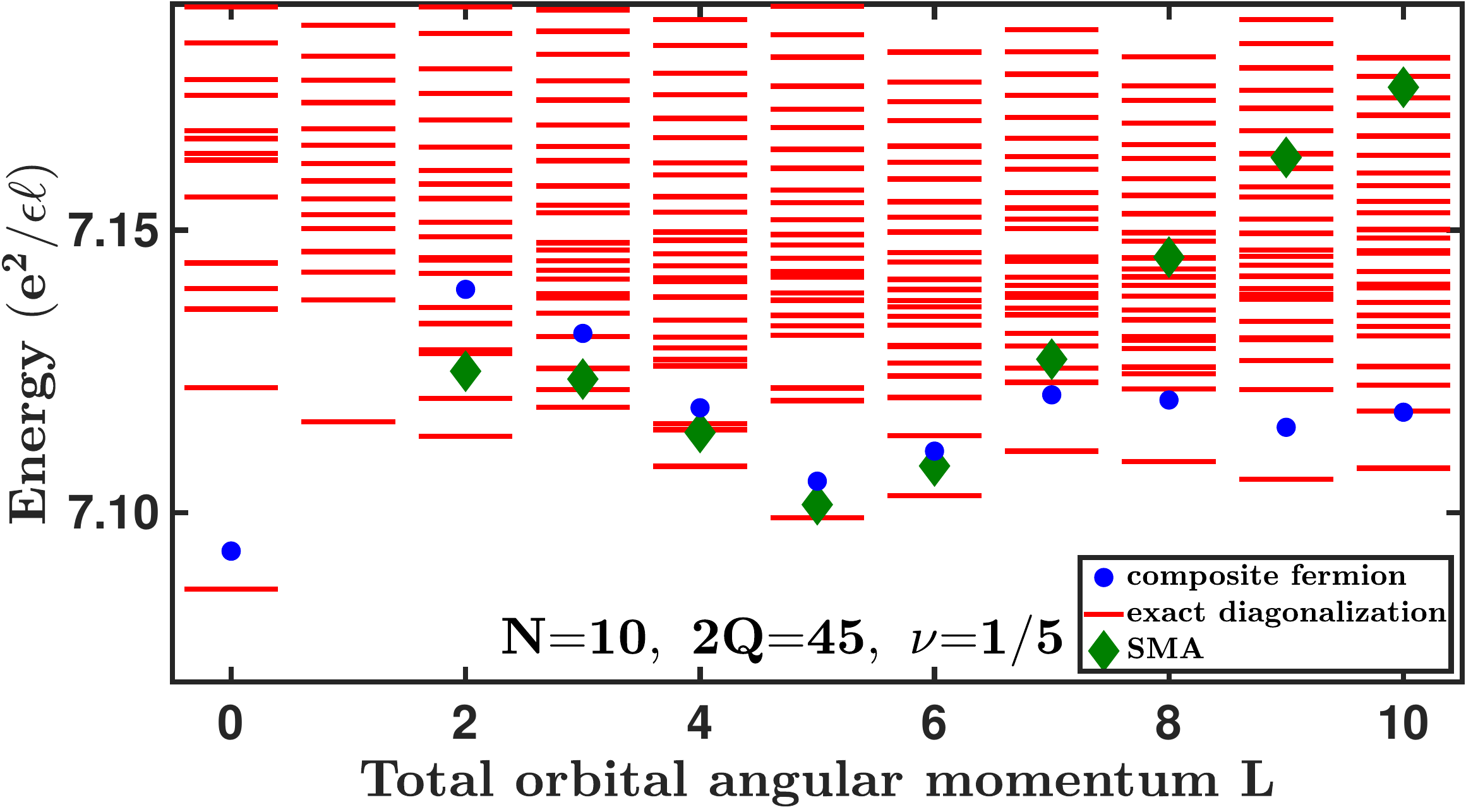}
	\includegraphics[width=0.32\linewidth]{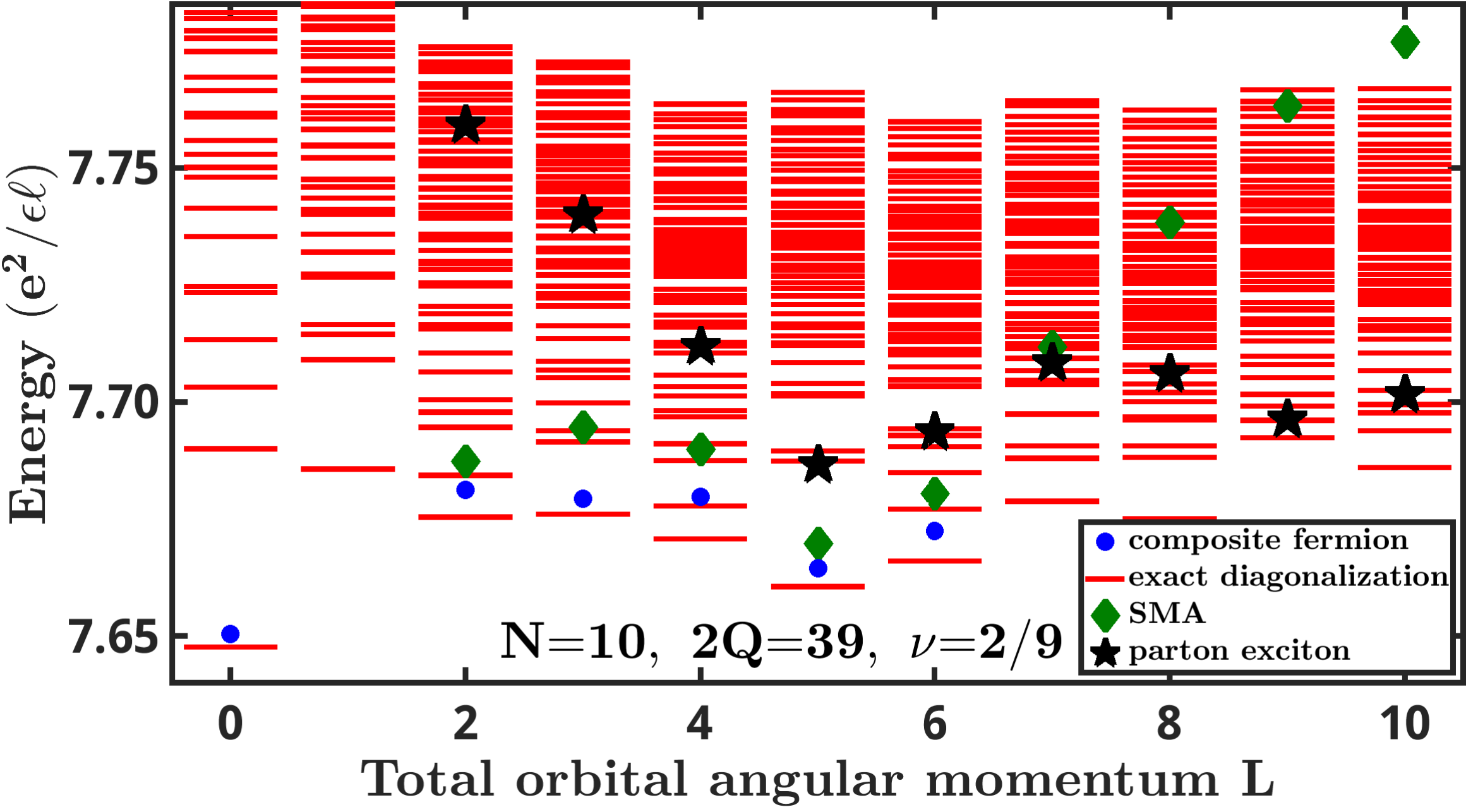}
	\includegraphics[width=0.32\linewidth]{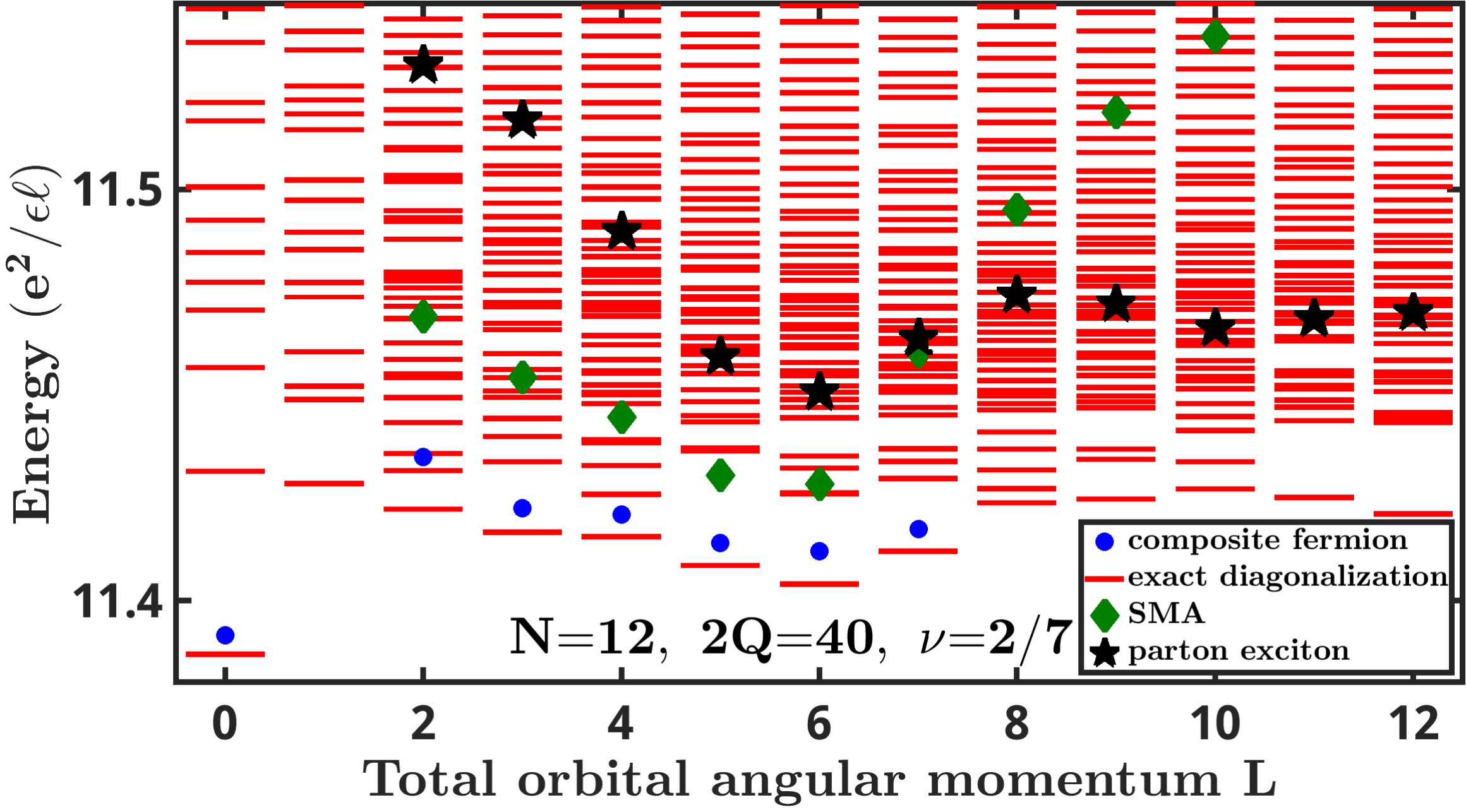}
	\caption{\label{fig: GMP_Laughlin_Jain_CFE_LLL_Coulomb_SMA_on_Coulomb} Same as Fig. 3 of the main text but with the GMP energy computed using the exact LLL Coulomb ground state.
	}
\end{figure*}

In Fig.~\ref{fig: GMP_Laughlin_Jain_CFE_LLL_Coulomb_SMA_on_Coulomb} we have presented results analogous to those shown in Fig. 3 of the main text, but with the GMP energy obtained using the exact LLL Coulomb ground state. (The primary CF and the parton excitons are the same as in Fig.~3 of the main text.) As expected, the energy of the GMP mode improves, but the results more or less mirror those shown in Fig. 3 with some minor quantitative differences. The largest difference is for $\nu{=}1/5$, where the GMP ansatz obtained using the LLL Coulomb ground state has significantly lower energy than the CF-exciton state in the long-wavelength limit. This follows from the fact that the Laughlin state~\cite{Laughlin83} does not provide as accurate a representation of the exact Coulomb ground state at 1/5 as it does at 1/3~\cite{Ambrumenil88,  Kusmierz18, Balram20a}.
 
\section {Effect of disorder broadening and resolution on the spectral function}
\label{sec: disorder_broadening_DSF}
\begin{figure}[t]
\centering
\includegraphics[width=0.85\textwidth]{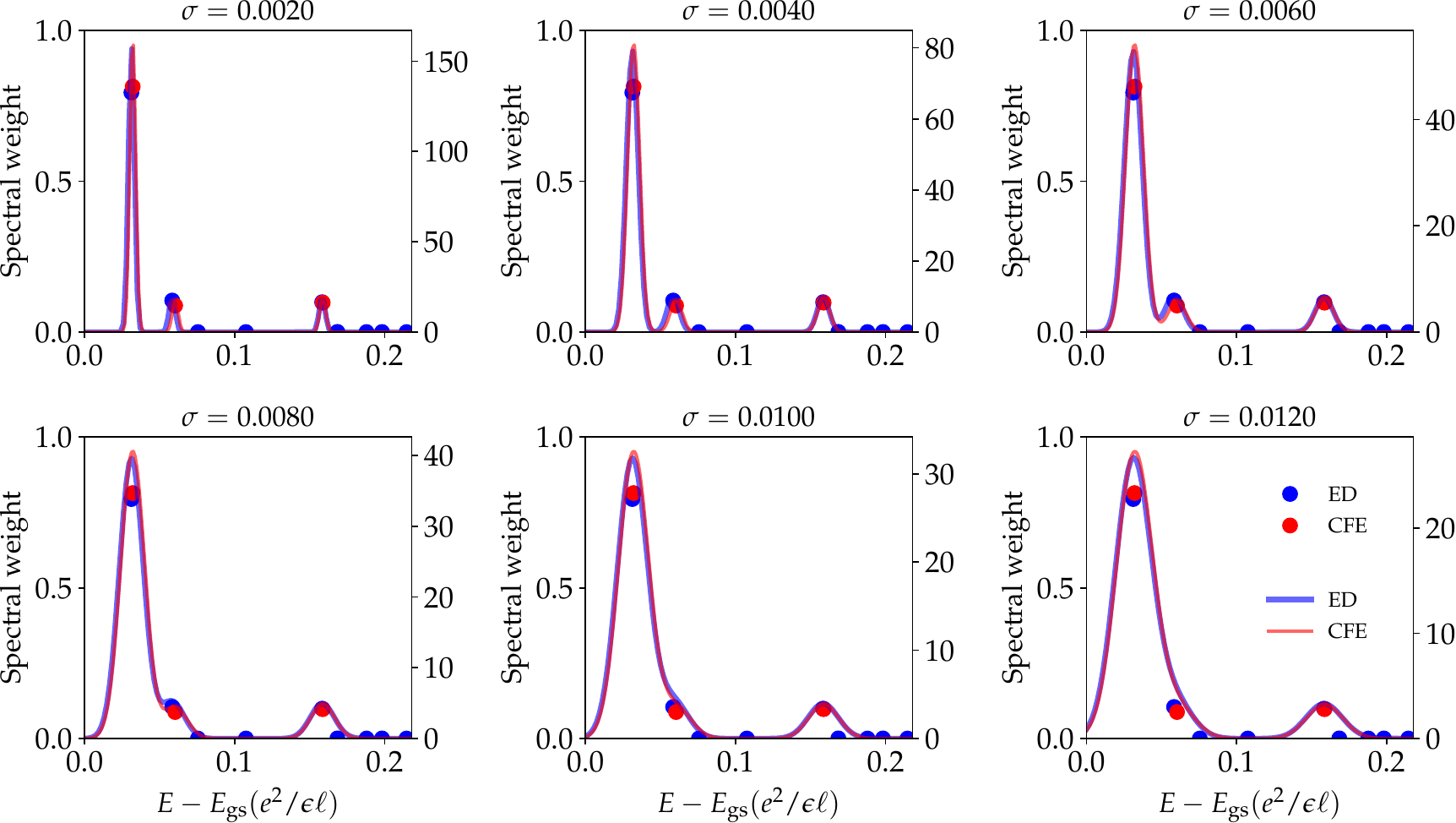}
\caption{\label{fig: SW296_width_palettes} Effect of bin-width on the spectral function of the GMP mode at $\nu{=}2/9$, $N{=}6$.  
As in Fig.~4 of the main text, the blue and red dots show spectral weights with the eigenstates obtained from diagonalization of Coulomb interaction in the full Hilbert space and CFE subspace respectively. The spectral function is estimated by replacing the delta functions in Eq.~(1) of the main text with normal distributions of width $\sigma$. Different panels show the spectral functions calculated using different choices of the $\sigma$.}
\end{figure}

The spectral function of the GMP mode (which forms the DSF) shown in Fig.~4 in the main text is estimated by replacing the delta functions appearing on the right-hand-side of Eq.~(1) in the main text with normal distributions of width $\sigma$ as follows
\begin{equation} 
S(\vec{q},E)\approx
\sum_\alpha |\langle \Psi_{\nu}^\alpha|  \Psi^{\rm GMP}_{\vec{q}}\rangle|^2 \frac{1}{\sqrt{2\pi \sigma}}\exp\left[ \frac{(E_\alpha-E)^2}{2\sigma^2}  \right].
\end{equation}
Fig.~\ref{fig: SW296_width_palettes} shows the effect of $\sigma$ on the estimated spectral function at filling $\nu{=}2/9$ of $N{=}6$ particles.
The value of $\sigma$ may be thought of as arising from the disorder-broadening of the peaks or the resolution of the measurement. The spectral function shows two prominent peaks for low resolution, i.e. for values of $\sigma$ larger than the energy level spacing within the $L{=}2$ angular momentum sector.

\section {Dynamical structure factor at other angular momenta}
\label{sec: DSF_L_geq2_2_5}
\begin{figure}[t]
\centering
\includegraphics[width=0.85\textwidth]{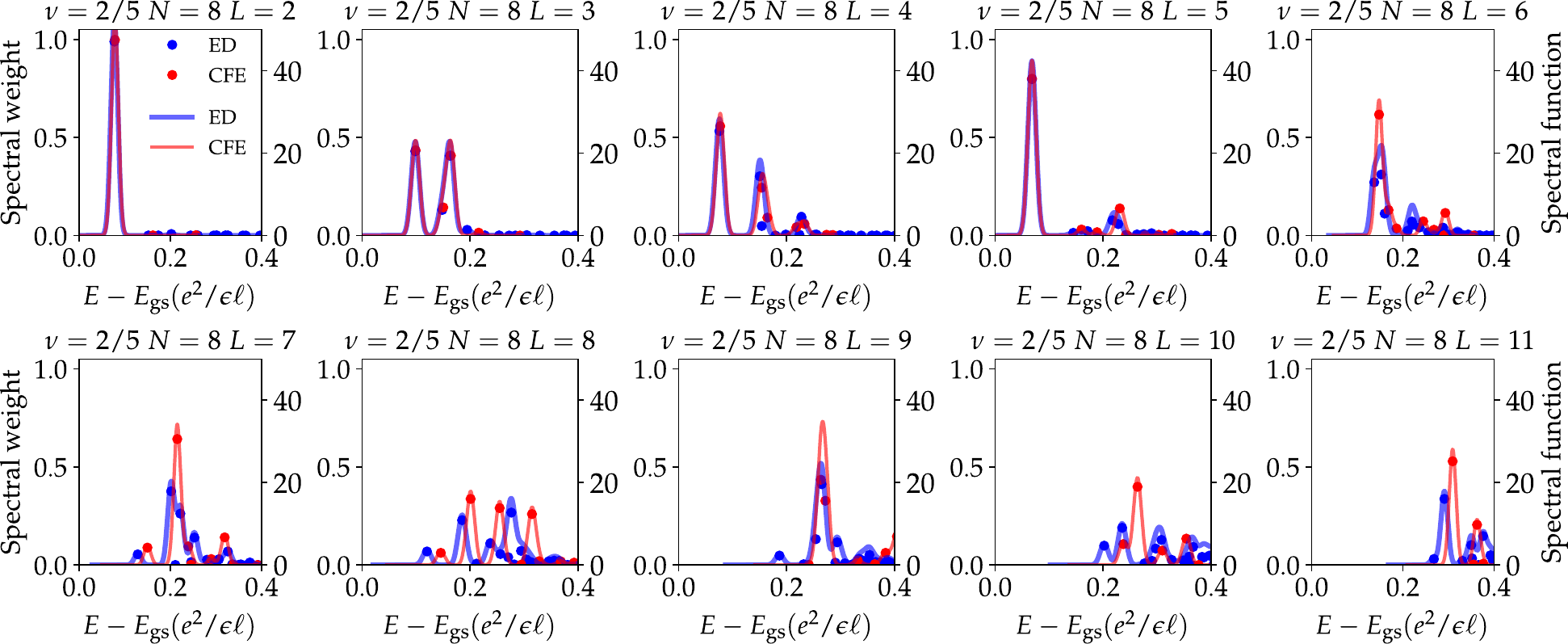}
\caption{\label{fig: SW258_full} Spectral weights (dots) and spectral functions (continuous lines) for the GMP mode on the $\nu{=}2/5$ state obtained from full Hilbert space ED calculations (blue) and direct projected CFE subspace (red) shown for the angular momenta sectors $2{\leq} L{\leq} 11$.}.  
\end{figure}

Figure~\ref{fig: SW258_full} shows the spectral weights and DSF for the GMP mode of the $\nu{=}2/5$ state for angular momentum sectors $L{=}2$ to $L{=}11$ for $N{=}8$ electrons. A single peak is visible at the lowest angular momentum $L{=}2$ as expected for a primary Jain sequence state, however, more peaks emerge at larger momenta. The DSF peaks appear at higher energies at larger momenta. The spectral functions forming obtained from calculations involving only the directed projected CFE space match the ones obtained by the calculations in the full Hilbert space till $L{\approx}9$ with the agreement being much better at lower angular momenta.

\end{document}